\documentclass[aps,prd,twocolumn,groupedaddress,showpacs,nofootinbib,amssymb,longbibliography]{revtex4-2}

\usepackage{graphicx}
\usepackage{amsmath,amssymb,amsfonts}
\usepackage{bm}
\usepackage{physics}
\usepackage{afterpage}
\usepackage{booktabs}
\usepackage{caption}
\usepackage{xcolor}

\usepackage{float}
\usepackage{url}
\usepackage[colorlinks=true, allcolors=blue]{hyperref}
\newcommand{\nside}{N_{\mathrm{side}}}

\newcommand{\Npix}{N_{\mathrm{pix}}}
\newcommand{\Pdet}{P_{\mathrm{det}}}
\newcommand{\Ndet}{N_{\mathrm{det}}}
\newcommand\be{\begin{equation}}
\newcommand\ee{\end{equation}}

\allowdisplaybreaks[4]
\begin{document}

\bibliographystyle{apsrev4-2}
\title{Orientation-dependent sensitivity of the all-sky
  all-frequency \\ gravitational-wave radiometer search to
  continuous-wave sources}
\author{Stavros Venikoudis$^{1}$}
\email{stavros.venikoudis@uclouvain.be}
\author{Giacomo Bruno$^{1}$}
\email{giacomo.bruno@uclouvain.be}
\affiliation{$^{1}$UCLouvain, Centre for Cosmology, Particle Physics and Phenomenology, CP3, Chemin du Cyclotron 2, 1348 Louvain-la-Neuve, Belgium}

\begin{abstract}
The all-sky all-frequency (ASAF) gravitational-wave radiometer search
cross-correlates the data of two detectors with equal weight on the
$+$ and $\times$ tensor polarizations. This choice is optimal for an
unpolarized stochastic gravitational-wave background, but not for
continuous-wave (CW) sources, whose polarization content depends on
the inclination angle $\iota$ and the polarization angle $\psi$. As a
result, unless the source is seen face-on, the search measures less
power than a face-on source with the same intrinsic amplitude would
produce, and underestimates its intrinsic strength. The ratio of the
two powers, which we call the projection factor $\mu^2(\iota,\psi)$, is known from
earlier radiometer studies. We show how $\mu^2$
enters the ASAF statistic and derive its exact dependence on $\psi$. We
also prove that its average over source orientations is exactly $2/5$,
independent of sky position, baseline geometry, noise weighting, and
frequency. We apply the formalism to four simulated pulsar signals at
four sky positions, with polarization ranging from circular to nearly
linear, injected into the LIGO detectors during the second observing
run (O2). A nearly edge-on injection ($\iota = 94.6^{\circ}$,
$\psi = -25.5^{\circ}$) is measured with a signal-to-noise ratio (SNR)
of $5.20$, below the detection threshold of $6.19$, while the same
signal seen face-on would have an SNR of $31.1 \pm 6.0$. Bayesian upper
limits on the intrinsic strain amplitude $h_0$, marginalized over
orientation and calibration uncertainty, are $2.7$ to $2.9$ times
larger than the face-on values. We then consider the $505$
sub-threshold candidates selected for follow-up investigation in the
first part of the fourth LIGO-Virgo-KAGRA observing run (O4a). For an
unknown orientation their upper limits increase by a factor of $2.5$ to
$2.9$, and each candidate acquires a range of upper limits over source
orientation in place of a single face-on value. If a candidate is a
real CW source at a typical orientation, a source of the same intrinsic
strength would reach the detection threshold only when inclined by less
than $34.7^{\circ}$ to $50.3^{\circ}$ from face-on. Accounting for
$\mu^2$ changes only the interpretation of the measured SNR and strain
amplitude. The measured SNR and the significance of every candidate are unchanged.
\end{abstract}

\maketitle
\section{Introduction}
\label{sec:introduction}
The detection of gravitational waves (GWs) by the LIGO-Virgo-KAGRA
(LVK)
collaboration~\cite{KAGRA:2013rdx,aLIGO:2020wna,LIGOScientific:2014pky,Virgo:2019juy,KAGRA:2020tym,Capote:2024rmo,Virgo:2026O4}
has opened new observational windows onto compact astrophysical
objects~\cite{LIGOScientific:2016aoc,LIGOScientific:2017vwq,LIGOScientific:2021qlt}.
Across the first three observing runs (O1--O3) and the first part of
the fourth (O4a), the network has cataloged over 200 compact binary
coalescences~\cite{LIGOScientific:2018mvr,LIGOScientific:2020ibl,KAGRA:2021vkt,LIGOScientific:2025slb,LIGOScientific:2025pvj}.
As detector sensitivity continues to improve, additional signal
classes are expected to become detectable. These include persistent
quasi-monochromatic emission from isolated, deformed neutron
stars~\cite{Riles:2017evm,Sieniawska:2019hmd,Wette:2023dom,Lasky:2015uia},
gravitational-wave bursts from transient
sources~\cite{LIGOScientific:2021hoh}, and stochastic
gravitational-wave backgrounds
(SGWBs)~\cite{Allen:1997ad,Romano:2016dpx,Christensen:2018iqi,Renzini:2022alw,vanRemortel:2022fkb,Virgo:2025aai, LIGOScientific:2025kry, LIGOScientific:2026mjf}.
An SGWB is produced by the overlap of many weak sources that cannot be
resolved individually. The angular distribution of this background
carries information about the nature of its sources. A cosmological
SGWB, such as one produced by
inflation~\cite{Grishchuk:1974ny,Starobinsky:1979ty}, phase
transitions~\cite{Caprini:2019egz,Hindmarsh:2020hop}, or cosmic
strings~\cite{Kibble:1976sj,Sarangi:2002yt,Auclair:2019wcv,Martin:1996ea},
is expected to be nearly isotropic. An astrophysical SGWB, produced by
compact objects concentrated along the Galactic plane and in galaxy
clusters~\cite{Jenkins:2018uac,Cusin:2018rsq,Mentasti:2023gmg,Rosado:2012bk,Zhu:2011bd,Talukder:2014eba,Buonanno:2004tp,Crocker:2015taa},
is expected to be anisotropic.

The standard technique for mapping this angular distribution is
gravitational-wave radiometry, which cross-correlates the data of
pairs of detectors~\cite{Allen:1997ad,Christensen:1992wi,Flanagan:1993ix}.
Two spatially separated detectors observe the same gravitational-wave
signal with a time delay that depends on the source direction. As the
Earth rotates, the delay changes, allowing different sky directions to
be distinguished during a sidereal
day~\cite{Ballmer:2005uw,Mitra:2007mc,Thrane:2009fp}. By accumulating
cross-correlated power as a function of both frequency and sky
position, one can build a map of the gravitational-wave
sky.

Several searches for anisotropic backgrounds have been performed based
on this approach~\cite{LIGOScientific:2016nwa, LIGOScientific:2019gaw, KAGRA:2021mth,KAGRA:2021rmt,LIGOScientific:2025bkz}. Narrowband radiometer analyses target selected sky positions and
frequency intervals~\cite{KAGRA:2021mth,LIGOScientific:2025bkz}.
Broadband radiometer analyses map the full sky for a background with
an assumed power-law spectrum, combining all frequencies into a single
estimate~\cite{KAGRA:2021mth,LIGOScientific:2025bkz}. Spherical
harmonic decompositions characterize the angular power spectrum of
backgrounds that are extended on the sky, also for an assumed spectral
shape~\cite{Thrane:2009fp,Ain:2018zvo,Renzini:2018vkx,Suresh:2020khz,LIGOScientific:2025bkz}.
Each method therefore fixes in advance either the sky directions or
the spectral shape. None estimates the power independently in every
sky direction and at every frequency.

The all-sky all-frequency (ASAF) radiometer search was developed to
fill this gap~\cite{KAGRA:2021rmt}. It computes the cross-correlated
power at every frequency bin and every sky pixel, without assuming a
source position or spectral shape. It is sensitive to persistent
narrowband signals. These include continuous-wave (CW) emission from
deformed rotating neutron stars and narrowband stochastic signals from
unresolved emitters along a given line of
sight~\cite{KAGRA:2021rmt,Agarwal:2021gvz}. Unlike matched-filter CW
searches, ASAF does not search over the parameters of the source. It
covers the whole sky and frequency band at a much lower computational
cost, at the price of sensitivity, and can therefore serve as a first
stage in hierarchical
searches~\cite{Tenorio:2021wmz,Riles:2022wwz,Knee:2023toa,Salvadore:2025lba}.
Because it does not model the phase evolution of the signal, it needs
no knowledge of the spin-down rate, orbital parameters, or glitch
history of the source, as long as the frequency of the signal stays
within one frequency bin of the search over the observing run. This
makes ASAF well suited to neutron stars with irregular spin
evolution~\cite{Lasky:2015uia} and to ultralight-boson condensates
around spinning black
holes~\cite{Arvanitaki:2009fg,Arvanitaki:2014wva,Riles:2022wwz}.

To date, no radiometer search has yielded a confirmed
detection~\cite{LIGOScientific:2016nwa,LIGOScientific:2019gaw,KAGRA:2021mth,KAGRA:2021rmt,LIGOScientific:2025bkz}.
The ASAF analysis of the O4a data reported a marginal outlier at
$92.8125$~Hz, where the all-sky map has a minimum signal-to-noise
ratio (SNR) of $-6.12$, just short of the two-sided threshold of
$-6.18$ used in that analysis, and a maximum of $5.2$ at a different sky position, which was
selected for follow-up. The follow-up checks were inconclusive, and
the feature will be monitored in future observing
runs~\cite{LIGOScientific:2025bkz}. In the absence of a detection,
results are reported as upper limits (ULs) on the GW strain amplitude,
assuming circular
polarization~\cite{Mitra:2007mc,Thrane:2009fp,KAGRA:2021mth,KAGRA:2021rmt,Agarwal:2021gvz,LIGOScientific:2025bkz}.
Since its introduction~\cite{KAGRA:2021rmt}, the ASAF analysis has
been studied further. The effect of correlated magnetic noise on its
cross-correlation statistic has been
quantified~\cite{Venikoudis:2024zmk}, a peak-finding algorithm has
been proposed to select fewer and less correlated candidates for
follow-up~\cite{Sharma:2025fyw}, and its sub-threshold candidates have
been followed up with dedicated CW
searches~\cite{Knee:2023toa,Knee:2026spotlight}. One issue that has
not yet been addressed is how much signal power the search loses when
its results are interpreted in terms of CW sources.

The origin of this loss is the equal weight that the cross-correlation
statistic gives to the two tensor polarizations, which assumes that
they carry equal
power~\cite{Mitra:2007mc,Thrane:2009fp,Nishizawa:2009bf}. This
assumption holds for an unpolarized SGWB. CW sources, however, have
emission amplitudes that depend on the inclination angle~$\iota$
between the neutron star's spin axis and the line of
sight~\cite{Jaranowski:1998qm,Dupuis:2005xz}. The two polarization
amplitudes are equal only for a face-on source ($\iota = 0$). At all
other inclinations, the equal-weight statistic measures less
power than a face-on source with the same amplitude would produce, and
the search underestimates the strength of the source. We call this
underestimate the orientation bias. In this work, we quantify it with
the projection factor $\mu^2(\iota,\psi)$, the ratio of the two
powers, where $\psi$ is the polarization angle. We apply the formalism to four pulsar hardware injections
(PHWIs)~\cite{Biwer:2016oyg}, simulated pulsar signals physically
injected into the LIGO detectors during the second observing run (O2),
and to the $505$ sub-threshold candidates of the O4a ASAF analysis
that were selected for follow-up.

Matched-filter CW pipelines account for inclination dependence by
searching over polarization
parameters~\cite{Jaranowski:1998qm,Dreissigacker:2018afk,Wette:2023dom},
but the current ASAF search does not, since it uses a single template
for all orientations. Modeled cross-correlation searches likewise
build the $(\iota,\psi)$ dependence into their
filters~\cite{Dhurandhar:2007vb,Whelan:2015dha}. For the SGWB,
\citet{Boybeyi:2026} have recently shown that maps that estimate only
the total power in each direction are biased when the sky is
polarized. The reason is that the detector network does not in
general respond to the intensity and to the polarization of a signal
in independent ways, so polarized power leaks into the intensity
estimate. They remove the bias with a maximum-likelihood analysis of
all four Stokes parameters, applied to O3 data. A CW source is a
point-like and fully polarized example of the same effect, and the
bias studied here is its consequence for the equal-weight ASAF
statistic.

The orientation factor that quantifies this bias is not itself new.
Averaged over the polarization angle, it falls from unity for a
face-on source to one eighth for an edge-on one, with an isotropic
average of $2/5$~\cite{Jaranowski:1998qm} (Sec.~\ref{sec:mu2}). Its
effect on radiometer searches was derived in
Refs.~\cite{Messenger:2010T1000195,Messenger:2015tga}, and narrowband
radiometer searches have used it to quote ULs marginalized over
orientation~\cite{Meyers:2018thesis,LIGOScientific:2025bkz}. What is
new here is threefold. First, we prove that the isotropic average of
$2/5$ holds exactly for the full ASAF statistic at any sky position.
Second, we derive the closed form of the residual dependence on the
polarization angle. It gives the least favorable orientation and the
largest detectable inclination at every sky position, and explains why
the marginalized ULs depend only on
declination~\cite{Messenger:2010T1000195}. Third, we propagate the
bias through the strain estimate, SNR, detection probability, and ULs
of the ASAF search. The ULs still assume circular polarization. With the bias included, each frequency-pixel pair has a range of ULs
over orientation instead of a single face-on value.

Section~\ref{sec:radiometer} summarizes the ASAF formalism and
analysis settings. Section~\ref{sec:cw_extension} derives
$\mu^2(\iota,\psi)$ for CW signals and propagates it through strain,
SNR, and detection probability, and Sec.~\ref{sec:marginalized_uls}
constructs orientation-marginalized ULs.
Sections~\ref{sec:orientation_results} and~\ref{sec:o4a_application}
apply the formalism to the four O2 PHWIs and to the $505$ O4a follow-up
candidates, and Sec.~\ref{sec:conclusions} summarizes the findings.

\section{The ASAF radiometer search}
\label{sec:radiometer}
In this section we summarize the ASAF formalism and analysis settings,
following
Refs.~\cite{Mitra:2007mc,Thrane:2009fp,Romano:2016dpx,KAGRA:2021mth,KAGRA:2021rmt,LIGOScientific:2025bkz}. As in the O3 and O4a ASAF
analyses~\cite{KAGRA:2021rmt,LIGOScientific:2025bkz}, we analyze
frequency bins of width $\Delta f = 1/32$~Hz. We divide the sky into
$\Npix = 12\,\nside^2$ equal-area HEALPix pixels~\cite{Gorski:2004by}
with $\nside = 16$. This gives $3\,072$ pixels of about
$13.4\,\mathrm{deg}^2$ each. The response of a baseline repeats every sidereal day. The data of a
full observing run can therefore be folded onto a single sidereal day
without loss of directional information. The folded data contain all
the data of the run, but cost only as much to analyze and store as a
single sidereal day~\cite{Thrane:2015aua,Ain:2015lea,Ain:2018zvo}.
They are analyzed with \texttt{PyStoch}~\cite{Ain:2018zvo}.

\subsection{Directional strain power}
\label{sec:omega_gw}
A directional search measures $\mathcal{P}(f,\hat{n})$, the
one-sided strain power spectral density per unit solid angle at
frequency~$f$ in sky direction~$\hat{n}$.
Its integral over the sky gives the total one-sided strain power
spectral density~\cite{Romano:2016dpx}. We expand it on a set of
basis functions $e_\mu(\hat{n})$,
\begin{equation}
  \mathcal{P}(f,\hat{n})
  = \sum_\mu \mathcal{P}_\mu(f)\;e_\mu(\hat{n})\,.
  \label{eq:P_basis}
\end{equation}
The pixel basis, $e_\mu(\hat{n}) = \delta^2(\hat{n}-\hat{n}_\mu)$,
targets point-like sources, and the spherical harmonic basis,
$e_\mu(\hat{n}) = Y_{\ell m}(\hat{n})$, targets spatially extended
emission~\cite{Thrane:2009fp,Renzini:2018vkx}. In this work we look for CW sources, which are point-like on the scale of a pixel. We therefore use the pixel basis.

\subsection{Cross-spectral density estimator}
\label{sec:csd}
We measure $\mathcal{P}(f,\hat{n})$ by cross-correlating the data of
two geographically separated detectors ($i = 1,2$), which form a
baseline~$I$. Noise that is uncorrelated between the two sites
averages out in the cross-correlation, while a GW signal common to
both accumulates~\cite{Allen:1997ad,Romano:2016dpx}. For each baseline
we construct the complex cross-spectral density (CSD)
estimator~\cite{KAGRA:2021mth,Allen:1997ad}
\begin{equation}
  C^I(t;f)
  = \frac{2}{T\;\overline{w_1 w_2}}\;
    \tilde{s}_1^{\ast}(t;f)\;\tilde{s}_2(t;f)\,,
  \label{eq:csd}
\end{equation}
where $\tilde{s}_i(t;f)$ is the short Fourier transform (SFT) of a
data segment of duration $T = 192$~s centered on time~$t$. The
segments are Hann windowed and overlapped by 50\%, and the factor
$\overline{w_1 w_2}$ corrects the estimator for the power removed by
the windows~\cite{Allen:1997ad,Romano:2016dpx}. The SFT resolution
$\delta f = 1/T \simeq 5.2$~mHz is coarse-grained to the analysis bin
width $\Delta f$ by averaging six adjacent bins.

For Gaussian, stationary noise that is uncorrelated between the sites,
the average of the CSD over noise and source realizations is
\begin{equation}
  \bigl\langle C^I(t;f)\bigr\rangle
  = \int d^2\hat{n}\;\gamma^I(t;f,\hat{n})\;\mathcal{P}(f,\hat{n})\,.
  \label{eq:csd_exp}
\end{equation}
A single CSD measurement is therefore a weighted integral of the GW
power over the whole sky. The weight, the directional response
function~$\gamma^I(t;f,\hat{n})$ defined below, maps the angular
distribution of the power onto the measured cross-correlation. It
changes as the Earth rotates, and this time dependence is what allows
the sky distribution of~$\mathcal{P}$ to be recovered.

\subsection{Overlap reduction function}
\label{sec:orf}
The directional response function $\gamma^I$ entering
Eq.~\eqref{eq:csd_exp} is called the overlap reduction function
(ORF)~\cite{Christensen:1992wi,Flanagan:1993ix,Finn:2008vh,Mitra:2007mc,Thrane:2009fp,Romano:2016dpx}.
It quantifies the combined response of the two detectors of a
baseline to a GW arriving from a given sky direction, at a given
frequency and time. In the pixel basis it reads
\begin{equation}
  \gamma^I(t;f,\hat{n})
  = \sum_{A=+,\times} F_1^A(\hat{n},t)\;F_2^A(\hat{n},t)\;
    e^{2\pi i f\,\hat{n}\cdot\Delta\vec{x}^I(t)/c}\,,
  \label{eq:orf}
\end{equation}
where $F_i^A(\hat{n},t)$ is the antenna pattern function of
detector~$i$ for polarization $A$, and $\Delta\vec{x}^I(t)$ is the
separation vector between the vertices of the two detectors. The
antenna patterns are evaluated in the small-antenna limit, in which the
arm length is negligible compared with the GW
wavelength~\cite{Romano:2016dpx}. They encode the orientation of the
arms of each detector with respect to the incoming wave, and the
exponential is the phase delay between the two sites. Both factors
depend on time through the rotation of the Earth and return to the
same values after one sidereal day, which is what allows the folding
described at the beginning of this section.

\subsection{Pixel-basis estimator}
\label{sec:pixel_basis}
At each frequency bin~$f$, the goal is to estimate the power
$\mathcal{P}_\mu(f)$ at every pixel~$\mu$. Assuming the CSD in each
time segment is Gaussian, the maximum-likelihood estimator
is~\cite{KAGRA:2021rmt,Ain:2018zvo,Mitra:2007mc}
\begin{equation}
  \hat{\mathcal{P}}_\mu(f)
  = \sum_{\mu'}\bigl(\Gamma^{-1}\bigr)_{\mu\mu'}(f)\;X_{\mu'}(f)\,,
  \label{eq:ML_estimator}
\end{equation}
where $X_\mu(f)$ is the narrowband dirty map and $\Gamma_{\mu\mu'}(f)$
is the narrowband Fisher information
matrix~\cite{KAGRA:2021rmt,Ain:2018zvo,Thrane:2009fp},
\begin{align}
  X_\mu(f)
  &\propto \Re\sum_{I,\,t}
    \frac{\gamma^{I\ast}_\mu(t;f)\;C^I(t;f)}
         {P_1(t;f)\;P_2(t;f)}\,,
  \label{eq:dirty_map}\\[6pt]
  \Gamma_{\mu\mu'}(f)
  &\propto \Re\sum_{I,\,t}
    \frac{\gamma^{I\ast}_\mu(t;f)\;\gamma^I_{\mu'}(t;f)}
         {P_1(t;f)\;P_2(t;f)}\,.
  \label{eq:fisher_pixel}
\end{align}
Here $\gamma^I_\mu(t;f) \equiv \gamma^I(t;f,\hat{n}_\mu)$, and
$P_{1,2}(t;f)$ are the one-sided noise power spectral densities of the
individual detectors. The sums run over all independent baselines $I$
and time segments $t$, and the real part combines the contributions of
positive and negative frequencies.

The off-diagonal elements of the Fisher matrix describe the
correlations between pixels caused by the finite angular resolution of
the baseline. Inverting the full matrix would remove them, but
directional blind spots make it highly
ill-conditioned~\cite{Mitra:2007mc,Panda:2019czb,Renzini:2018nee,Renzini:2019vmt,Agarwal:2021gvz,Xiao:2023hcl,Suresh:2020khz}.
The ASAF search therefore keeps only the diagonal
elements~\cite{KAGRA:2021rmt}. For a point-like source, whose power is
contained in a single pixel~$\hat{n}$, the expected dirty map at that
pixel is $\langle X_{\hat{n}}\rangle = \Gamma_{\hat{n}\hat{n}}\,
\mathcal{P}_{\hat{n}}$, so the diagonal estimate is unbiased
there~\cite{Agarwal:2021gvz}. The resulting point estimate and its
variance are
\begin{equation}
  \hat{\mathcal{P}}_{\hat{n}}(f)
  = \bigl[\Gamma_{\hat{n}\hat{n}}(f)\bigr]^{-1} X_{\hat{n}}(f)\,,
  \qquad
  \sigma_{\hat{n}}^2(f)
  = \bigl[\Gamma_{\hat{n}\hat{n}}(f)\bigr]^{-1}\,,
  \label{eq:point_estimate}
\end{equation}
with no summation over $\hat{n}$. The detection statistic is the
signal-to-noise ratio (SNR)
\begin{equation}
  \hat{\rho}_{\hat{n}}(f)
  \equiv \frac{\hat{\mathcal{P}}_{\hat{n}}(f)}
              {\sigma_{\hat{n}}(f)}\,.
  \label{eq:snr}
\end{equation}

\subsection{Effective strain and null-hypothesis testing}
\label{sec:heff}
For nearly monochromatic CW signals, the ASAF search reports the
measured cross-correlated power as an effective strain
amplitude~\cite{KAGRA:2021rmt,KAGRA:2021mth,LIGOScientific:2025bkz},
\begin{equation}
  h_{\mathrm{eff}}(f,\hat{n})
  = \sqrt{\hat{\mathcal{P}}_{\hat{n}}(f)\,\Delta f}\,.
  \label{eq:heff_def}
\end{equation}
This is the strain that a circularly polarized, face-on source would
need in order to produce the observed power. The conversion between
radiometer quantities and the CW strain amplitude was developed for
the radiometer searches directed at Scorpius
X-1~\cite{Messenger:2010T1000195,Messenger:2015tga,Meyers:2018thesis}.
It equals the intrinsic amplitude $h_0$ only if the signal is
circularly polarized and confined to a single frequency
bin~\cite{KAGRA:2021rmt,KAGRA:2021mth,LIGOScientific:2025bkz}. The
second condition fails at high frequency. The annual Doppler
modulation of the source frequency by the motion of the Earth, of
order $\pm 10^{-4} f$, exceeds one $1/32$~Hz bin above roughly
$160$~Hz and spreads a line over neighboring
bins~\cite{LIGOScientific:2025bkz}. Intrinsic spin-down does the same
once $|\dot{f}| \gtrsim \Delta f/T_{\mathrm{obs}} \simeq
1.4\times10^{-9}$~Hz\,s$^{-1}$ for an O2-length run, far above the
rates of the injections used here. This spectral spreading is
independent of the source orientation. We do not correct for it,
but estimate its effect on the injections in
Sec.~\ref{sec:test_cases}. The first condition is the subject of this
paper. A CW source has a polarization state set by its orientation, and the
relation between $h_{\mathrm{eff}}$ and $h_0$ is then given by the
projection factor~$\mu^2$ derived in Sec.~\ref{sec:mu2}.

Under the null hypothesis the data contain only noise, for which
$\hat{\rho}$ would follow a zero-mean, unit-variance Gaussian
distribution. Detector data, however, contain non-Gaussian artifacts
such as narrowband spectral lines and
glitches~\cite{LSC:2018vzm,LIGO:2021ppb,LIGO:2024kkz,O4LIGODetector:2026okh},
so the null distribution is obtained with the random time-shift
method~\cite{KAGRA:2021rmt}. The run is divided into segments of a few
thousand seconds, and in each segment the data of one detector are
shifted by a random delay of $1$ to $2$~s. The delay is much longer
than the $10$~ms light travel time between the sites and differs
between segments, so any persistent GW signal loses its coherence and
only the noise background remains~\cite{KAGRA:2021rmt,Thrane:2009fp}.
The zero-lag data, without the time shift, are then tested against
this distribution.

The highest zero-lag SNR across all frequency-pixel pairs is assigned
a local $p$-value $p_1$, which is converted into a global one with
\v{S}id\'{a}k's correction~\cite{sidak1967},
$p_G = 1 - (1-p_1)^{N_{\mathrm{trials}}}$, to account for the many
pairs tested. The null hypothesis is rejected if $p_G < 0.05$, which
corresponds to the SNR threshold
\begin{equation}
  \rho_{\mathrm{thr}}
  = \sqrt{2}\;\mathrm{erfc}^{-1}\!\bigl(2\,p_1\bigr)\,,
  \qquad
  p_1 = 1 - (1-p_G)^{1/N_{\mathrm{trials}}}\,.
  \label{eq:threshold}
\end{equation}
With $\nside = 16$ ($\Npix = 3\,072$) and
$N_{\mathrm{freq}} = 54\,593$ frequency bins,
$N_{\mathrm{trials}} \approx 1.68 \times 10^8$, the total number of
frequency-pixel pairs. Setting $p_G = 0.05$ gives
$\rho_{\mathrm{thr}} = 6.19$, which we adopt throughout. Sub-threshold
candidates are identified by retaining the highest-SNR pixel at each
frequency. Each is compared to the time-shifted distribution within
the same $10$~Hz sub-band, and those above the 99th percentile are
selected for further
investigation~\cite{KAGRA:2021rmt,LIGOScientific:2025bkz,Knee:2023toa}.

\section{Continuous-wave sources and the orientation bias factor}
\label{sec:cw_extension}
The ASAF statistic of Sec.~\ref{sec:radiometer} is built on the SGWB
formalism, which assumes an unpolarized
signal~\cite{Allen:1997ad,Romano:2016dpx}. A CW source differs in two
ways. It is a quasi-monochromatic, coherent signal with a definite
phase evolution~\cite{Jaranowski:1998qm,Dhurandhar:2007vb,Whelan:2015dha},
and it emits in a polarization state fixed by its inclination and
polarization angle. The first difference determines the response of a
baseline to the source, and the second produces the orientation bias.
In Sec.~\ref{sec:cw_signal} we define the CW signal model, in
Sec.~\ref{sec:cw_crosscorr} we give the baseline response, in
Sec.~\ref{sec:mu2} we define the projection factor $\mu^2(\iota,\psi)$
and derive its orientation dependence, and in
Sec.~\ref{sec:orientation_chain} we propagate it to the quantities
reported by the search.

\subsection{Continuous-wave signal model}
\label{sec:cw_signal}
We consider a deformed rotating neutron star at a distance~$D$ from
Earth with equatorial ellipticity~$\epsilon$. Its time-varying
quadrupole moment emits gravitational radiation at twice the rotation
frequency, $f_0 = 2f_{\mathrm{rot}}$, with intrinsic strain
amplitude~\cite{Jaranowski:1998qm,Riles:2017evm,Lasky:2015uia,Sieniawska:2019hmd}
\begin{equation}
  h_0 = \frac{4\pi^2 G\,I_{zz}\,\epsilon\,f_0^2}{c^4\,D}\,,
  \label{eq:h0_def}
\end{equation}
where $I_{zz}$ is the moment of inertia about the rotation axis. The
source orientation is described by the inclination~$\iota$ and the
polarization angle~$\psi$, which set the relative amplitudes of the
two GW
polarizations~\cite{Jaranowski:1998qm,Dupuis:2005xz,Dreissigacker:2018afk}
\begin{equation}
\begin{aligned}
  A_+      &= \tfrac{1}{2}(1+\cos^2\iota)\,h_0
            &\equiv \mathcal{A}_+(\iota)\,h_0\,, \\
  A_\times &= \cos\iota\,h_0
            &\equiv \mathcal{A}_\times(\iota)\,h_0\,.
\end{aligned}
\label{eq:Apm_h0}
\end{equation}
A face-on source ($\iota = 0$) has
$\mathcal{A}_+ = \mathcal{A}_\times = 1$ and emits circularly
polarized radiation. An edge-on source ($\iota = \pi/2$) has
$\mathcal{A}_\times = 0$ and $\mathcal{A}_+ = 1/2$, producing linearly
polarized emission. Each detector responds to the incoming GW through
its antenna pattern functions, which separate the sky-position
dependence from the polarization-angle
dependence~\cite{Jaranowski:1998qm},
\begin{align}
  F_i^+(\hat{n},t;\psi)
    &= a_i(\hat{n},t)\cos 2\psi - b_i(\hat{n},t)\sin 2\psi\,,
  \label{eq:Fplus}\\
  F_i^\times(\hat{n},t;\psi)
    &= a_i(\hat{n},t)\sin 2\psi + b_i(\hat{n},t)\cos 2\psi\,,
  \label{eq:Fcross}
\end{align}
where the beam-pattern coefficients depend only on the source
direction,
\begin{align}
  a_i(\hat{n},t)
    &= \tfrac{1}{2}(1+\cos^2\theta)\cos 2\phi\,,
  \label{eq:ai}\\
  b_i(\hat{n},t)
    &= \cos\theta\sin 2\phi\,.
  \label{eq:bi}
\end{align}
Here $\theta$ and $\phi$ are the polar and azimuthal angles of the
source direction~$\hat{n}$ in the frame of detector~$i$ at time~$t$,
which change over time as the Earth rotates. The detector tensor is
$d_i^{ab} = (\hat{u}^a\hat{u}^b - \hat{v}^a\hat{v}^b)/2$, so that
$d_i^{ab}h_{ab}$ is the fractional change in arm length and
$|F_i^A| \leq 1$~\cite{Anderson:2000yy}.

\subsection{Cross-correlation response for a CW source}
\label{sec:cw_crosscorr}
The response of a baseline to a CW source was derived in
Refs.~\cite{Dhurandhar:2007vb,Whelan:2015dha}. We summarize it here
and give the derivation in Appendix~\ref{app:cw_response}. Let $I_1$
and $I_2$ denote the two detectors of baseline $I$, so that
$F^A_{I_i}$ is Eq.~\eqref{eq:Fplus} or~\eqref{eq:Fcross} evaluated in
the frame of that detector. The strain induced in detector $I_i$ by
the CW source of Sec.~\ref{sec:cw_signal} is
\begin{equation}
\begin{split}
  h_{I_i}(t) = {}& A_+\,F^+_{I_i}(\hat{n},t;\psi)\,
                   \cos\Phi_{I_i}(t;\Lambda)\\
             + {}& A_\times\,F^\times_{I_i}(\hat{n},t;\psi)\,
                   \sin\Phi_{I_i}(t;\Lambda)\,.
\end{split}
\label{eq:h_detector}
\end{equation}
The phase measured at the detector, $\Phi_{I_i}(t;\Lambda)$, combines
an intrinsic evolution set by $f_0$ and $\dot{f}_0$ with a Doppler
term fixed by the detector position $\vec{x}_{I_i}(t)$ relative to the
solar system barycenter~\cite{Jaranowski:1998qm,Wette:2023dom}. For an
isolated source, the signal is described by the amplitude $h_0$ and
the parameters
$\Lambda = \{\hat{n}, f_0, \dot{f}_0, \iota, \psi, \Phi_0\}$, with
$\Phi_0$ the initial phase. A source in a binary also requires the
orbital parameters.

Within an SFT of duration $T = 192$~s the signal is monochromatic to a
good approximation, so its SFT consists of two narrow lines at
$\pm f_0(t)$ (Appendix~\ref{app:cw_response}). At the
positive-frequency line, the product of the two signal SFTs entering
the CSD of Eq.~\eqref{eq:csd} is
\begin{equation}
  \tilde{h}^{*}_{I_1}(t;f)\,\tilde{h}_{I_2}(t;f)
  = \frac{1}{4}\,h_0^2\,G_I(t;\Lambda)\,
    \delta_T^2\!\bigl(f-f_0(t)\bigr)\,,
  \label{eq:signal_crosspower}
\end{equation}
where $\delta_T$ is the finite-time Dirac delta of
Eq.~\eqref{eq:deltaT}, and the complex baseline response function
is~\cite{Dhurandhar:2007vb,Whelan:2015dha}
\begin{equation}
\begin{split}
  G_I(t;\Lambda)
  = \bigl[\mathcal{A}_+\,F^+_{I_1}(\hat{n},t;\psi)
    &+ i\,\mathcal{A}_\times\,F^\times_{I_1}(\hat{n},t;\psi)\bigr]\\
    \times\bigl[\mathcal{A}_+\,F^+_{I_2}(\hat{n},t;\psi)
    &- i\,\mathcal{A}_\times\,F^\times_{I_2}(\hat{n},t;\psi)\bigr]
    \,e^{i\Delta\Phi_I(t)}\,.
\end{split}
  \label{eq:GI_def}
\end{equation}
The amplitude factors out as $h_0^2$, so $G_I$ depends only on the
geometry and the orientation. The source phase cancels between the two
detectors, and
$\Delta\Phi_I(t) = 2\pi f\,\hat{n}\cdot\Delta\vec{x}^I(t)/c$ is the
geometric delay of the ORF of Eq.~\eqref{eq:orf}.

We now fix the baseline and the sky direction, write $F^A_{1,2}$ for
$F^A_{I_{1,2}}(\hat{n},t;\psi)$, and carry the remaining time
dependence in the subscript of $G_t$. We write $\gamma_t$ for the
point-source ORF of Eq.~\eqref{eq:orf} reduced the same way. The
phase-delay factor $e^{i\Delta\Phi_I(t)}$ is common to $G_t$ and
$\gamma_t$, so it cancels in the projection factor defined below and
we drop it from both. Then $\gamma_t = F_1^+F_2^+ +
F_1^\times F_2^\times$ is real, and the instantaneous response is
\begin{equation}
  G_t(\iota,\psi)
  = \bigl[\mathcal{A}_+ F_1^+ + i\,\mathcal{A}_\times F_1^\times\bigr]
    \bigl[\mathcal{A}_+ F_2^+ - i\,\mathcal{A}_\times F_2^\times\bigr]\,.
  \label{eq:Gt}
\end{equation}
Its real part, which is the part retained by the estimator of
Sec.~\ref{sec:pixel_basis}, is
\begin{equation}
  \mathrm{Re}\,G_t
  = \mathcal{A}_+^2\,F_1^+F_2^+
  + \mathcal{A}_\times^2\,F_1^\times F_2^\times\,.
  \label{eq:ReGt}
\end{equation}
The cross terms, proportional to $\mathcal{A}_+\mathcal{A}_\times$,
are purely imaginary and drop out. The ORF weights the two
polarization products equally, whereas $\mathrm{Re}\,G_t$ weights them
by $\mathcal{A}_+^2$ and $\mathcal{A}_\times^2$. The two coincide only
if $\mathcal{A}_+^2 = \mathcal{A}_\times^2$, that is, for
$\cos\iota = \pm 1$.

\subsection{The projection factor \texorpdfstring{$\mu^2$}{mu2}}
\label{sec:mu2}
For a CW source with known parameters, the optimal estimator of
$\mathcal{H} = h_0^2$ uses $G_I$ in place of $\gamma^I$ as the
template~\cite{Dhurandhar:2007vb,Whelan:2015dha},
$\hat{\mathcal{H}} = \Gamma_{\mathrm{CW}}^{-1} X_{\mathrm{CW}}$, with
\begin{align}
  X_{\mathrm{CW}}(f) &\propto \mathrm{Re}\sum_{I,\,t}
    \frac{C^I(t;f)\,G^*_I(t;\Lambda)\,
      \delta_T^2\bigl(f-f_0(t)\bigr)}
         {P_{I_1}(t;f)\,P_{I_2}(t;f)}\,,
  \label{eq:X_cw}\\[4pt]
  \Gamma_{\mathrm{CW}}(f) &\propto \sum_{I,\,t}
    \frac{|G_I(t;\Lambda)|^2\,
      \delta_T^4\bigl(f-f_0(t)\bigr)}
         {P_{I_1}(t;f)\,P_{I_2}(t;f)}\,,
  \label{eq:Gamma_cw}
\end{align}
and variance $\Gamma_{\mathrm{CW}}^{-1}$. This matched filter recovers
$h_0^2$ at any orientation, but requires a bank of templates over
$(\hat{n}, f_0, \dot{f}_0, \iota, \psi)$. The ASAF search uses the
orientation-independent template $\gamma^I$ instead, with a single
template per sky pixel and frequency bin.

Applying the pixel estimator of Eq.~\eqref{eq:point_estimate} to the
CW signal gives the projection factor, called the bias factor in
Ref.~\cite{Meyers:2018thesis},
\begin{equation}
  \mu^2(\iota,\psi)
  = \frac{\sum_t w_t\,\gamma_t\,\mathrm{Re}\,G_t(\iota,\psi)}
         {\sum_t w_t\,\gamma_t^2}\,,
  \qquad
  w_t = \frac{1}{P_1(t;f)\,P_2(t;f)}\,.
  \label{eq:mu2}
\end{equation}
The numerator is the dirty map of Eq.~\eqref{eq:dirty_map} evaluated
on the CW signal, which is proportional to $G_t$, and the denominator
is the Fisher element of Eq.~\eqref{eq:fisher_pixel}. With the
normalization of Sec.~\ref{sec:pixel_basis}, which absorbs
the factor $1/4$ and the factor $\delta_T^2$, a line confined to a
single bin gives $\hat{\mathcal{P}} = \mu^2 h_0^2/\Delta f$, so that
$h_{\mathrm{eff}} = \sqrt{\mu^2}\,h_0$ in Eq.~\eqref{eq:heff_def}. The
projection factor is therefore the ratio of the effective to the
intrinsic strain power, $\mu^2 = h_{\mathrm{eff}}^2/h_0^2$. For a
face-on source $\mathrm{Re}\,G_t = \gamma_t$ and $\mu^2 = 1$ exactly.
Since $\mathrm{Re}\,G_t$ depends on $\iota$ only through
$\mathcal{A}_+^2$ and $\mathcal{A}_\times^2$, which are even in
$\cos\iota$, $\mu^2(\pi-\iota,\psi) = \mu^2(\iota,\psi)$, and we
restrict to $\cos\iota \in [0,1]$. The results below hold for any
positive weights. In our numerical evaluations we take $w_t$ uniform
over the sidereal day. For folded data this is a good approximation,
because the daily variations of the detector noise follow solar rather
than sidereal time and average out over a run of many months.

A simple limit illustrates the orientation dependence. For co-aligned
detectors with the source at the zenith ($a_i = 1$, $b_i = 0$), the
antenna patterns reduce to $F_i^+ = \cos 2\psi$ and
$F_i^\times = \sin 2\psi$, so $\gamma_t = 1$ and Eq.~\eqref{eq:ReGt}
gives
\begin{equation}
  \mu^2\big|_{\mathrm{zen}}
  = \mathcal{A}_+^2\cos^2 2\psi
  + \mathcal{A}_\times^2\sin^2 2\psi\,.
  \label{eq:mu2_iso_explicit}
\end{equation}
Averaging over $\psi$, uniformly distributed over one period
$[-\pi/4,\pi/4]$, gives
\begin{equation}
  \langle\mu^2\rangle_\psi
  = \frac{1+6\cos^2\iota+\cos^4\iota}{8}\,,
  \label{eq:mu2_psi_avg}
\end{equation}
the orientation factor commonly used in CW sensitivity
estimates~\cite{Jaranowski:1998qm}, which falls from $1$ at
$\iota = 0$ to $1/8$ at $\iota = \pi/2$. Averaging further over
$\cos\iota$ uniformly distributed in $[0,1]$, which corresponds to
isotropically oriented sources, yields
\begin{equation}
  \boxed{\;\langle\mu^2\rangle_{\iota,\psi}
  = \frac{2}{5} = 0.40\,.\;}
  \label{eq:mu2_isotropic}
\end{equation}
For randomly oriented sources, $h_{\mathrm{eff}}^2$ is therefore on
average $40\%$ of $h_0^2$, and $h_{\mathrm{eff}}$ underestimates $h_0$
by a factor $\sqrt{5/2} \approx 1.58$.

Using Eqs.~\eqref{eq:Fplus} and~\eqref{eq:Fcross}, the two
polarization products can be written as
(Appendix~\ref{app:closedform})
\begin{align}
  F_1^+F_2^+ + F_1^\times F_2^\times
    &= a_1a_2 + b_1b_2 \equiv A_t\,,
  \label{eq:idsum}\\
  F_1^+F_2^+ - F_1^\times F_2^\times
    &= C^{(1)}_t\cos 4\psi - C^{(2)}_t\sin 4\psi\,,
  \label{eq:iddiff}
\end{align}
with $C^{(1)}_t = a_1a_2 - b_1b_2$ and $C^{(2)}_t = a_1b_2 + a_2b_1$.
The sum is $\gamma_t$ and carries no $\psi$, and the difference carries
all of it. Summing over the sidereal day with the weights $w_t$ leaves
three numbers,
\begin{equation}
\begin{gathered}
  D = \sum_t w_t A_t^2\,,
  \qquad
  C^{(1)} = \sum_t w_t A_t C^{(1)}_t\,,\\
  C^{(2)} = \sum_t w_t A_t C^{(2)}_t\,,
\end{gathered}
  \label{eq:timesums}
\end{equation}
which define the depth and phase of the $\psi$ modulation,
\begin{equation}
  \varepsilon
  = \frac{\sqrt{(C^{(1)})^2 + (C^{(2)})^2}}{D}\,,
  \qquad
  \delta = \operatorname{atan2}\bigl(C^{(2)}, C^{(1)}\bigr)\,.
  \label{eq:epsdelta}
\end{equation}
Inserting Eqs.~\eqref{eq:idsum} and~\eqref{eq:iddiff} into
Eq.~\eqref{eq:mu2} gives the exact closed form
\begin{equation}
  \boxed{\;\mu^2(\iota,\psi)
  = \frac{\mathcal{A}_+^2+\mathcal{A}_\times^2}{2}
  + \frac{\mathcal{A}_+^2-\mathcal{A}_\times^2}{2}\,
    \varepsilon\,\cos\!\bigl(4\psi+\delta\bigr)\,.\;}
  \label{eq:theorem}
\end{equation}
The first term depends only on the inclination. The second term carries all the dependence on
$\psi$, as a single harmonic in $4\psi$. The parameters $\varepsilon$
and $\delta$ are not free. They are fixed by the baseline geometry,
and with uniform weights they are independent of frequency and depend
only on the declination of the source (Appendix~\ref{app:closedform}). At edge-on
inclination $\mu^2 = (1+\varepsilon\cos(4\psi+\delta))/8$, so
$\varepsilon$ is the relative depth of the $\psi$ modulation and
$\delta$ locates its maximum. The zenith limit of
Eq.~\eqref{eq:mu2_iso_explicit} corresponds to $\varepsilon = 1$,
$\delta = 0$. The modulation is not negligible. Its depth reaches $\varepsilon = 0.39$ at declinations near $\pm 19.5^\circ$.

Equation~\eqref{eq:theorem} has three exact consequences. First,
Eq.~\eqref{eq:mu2_psi_avg} holds for the full time-dependent statistic
at any sky position and for any noise weighting, so the zenith limit is
not needed. Second, integrating over $\cos\iota$ reproduces
Eq.~\eqref{eq:mu2_isotropic} independent of geometry, weighting, and
frequency. Third, provided $\varepsilon \leq 1$, the global minimum
lies at edge-on inclination,
\begin{equation}
  \mu^2_{\min} = \frac{1-\varepsilon}{8}
  \quad\mathrm{at}\quad
  \psi^* = \frac{\pi-\delta}{4}\ \ (\mathrm{mod}\ \pi/2)\,,
  \label{eq:mumin}
\end{equation}
which tends to $1/8$ near the celestial poles, where
$\varepsilon \to 0$ because the rotation of the Earth sweeps the
polarization angle uniformly over a day. The bound
$\varepsilon \leq 1$ is verified numerically for the O2 H1-L1 baseline
in Appendix~\ref{app:closedform}.

Finally, Eq.~\eqref{eq:theorem} fixes the largest inclination at which
a source remains detectable. Requiring
$\max_\psi \mu^2 = \rho_{\mathrm{thr}}/\rho_{\mathrm{ref}}$ gives
\begin{equation}
  (1+\varepsilon)\,u^2 + (6-2\varepsilon)\,u
  + (1+\varepsilon-8m) = 0\,,
  \label{eq:imaxexists}
\end{equation}
where $u = \cos^2\iota$ and
$m = \rho_{\mathrm{thr}}/\rho_{\mathrm{ref}}$. Its root $u \in [0,1]$
gives the largest inclination detectable at the most favorable $\psi$.
At $\varepsilon = 0$ it reduces to $\langle\mu^2\rangle_\psi = m$,
with $\langle\mu^2\rangle_\psi$ of Eq.~\eqref{eq:mu2_psi_avg}. Table~\ref{tab:sgwb_vs_cw}
summarizes the comparison between the stochastic and CW formalisms.

\onecolumngrid
\begin{center}
\begin{minipage}{\textwidth}
\captionof{table}{Comparison between the stochastic-background and
  continuous-wave cross-correlation formalisms. The equal-weight
  statistic is designed for the SGWB case. Applied to a CW source, it
  returns $h_{\mathrm{eff}}^2 = \mu^2 h_0^2$.}
\label{tab:sgwb_vs_cw}
\begin{ruledtabular}
\renewcommand{\arraystretch}{1.25}
\begin{tabular}{lll}
  Property & SGWB (assumed) & CW (actual) \\
  \colrule
  Signal type
    & random field
    & deterministic waveform [Eq.~\eqref{eq:h_detector}] \\
  Polarization content
    & equal, uncorrelated power in $+$ and $\times$
    & fixed by $(\iota,\psi)$ [Eq.~\eqref{eq:Apm_h0}] \\
  Baseline response
    & ORF $\gamma^{I}$ [Eq.~\eqref{eq:orf}]
    & complex $G_{I}$ [Eq.~\eqref{eq:GI_def}] \\
  Optimal filter
    & $\gamma^{I}$
    & $G_{I}$ [Eq.~\eqref{eq:X_cw}] \\
  Filter used by the radiometer
    & $\gamma^{I}$
    & $\gamma^{I}$ (not matched to $G_I$) \\
  Polarization weighting
    & equal weight for $+$ and $\times$
    & $\mathcal{A}_{+}^{2}$ and $\mathcal{A}_{\times}^{2}$ \\
  Recovered quantity
    & angular power $\mathcal{P}(f,\hat{n})$
    & $\mu^{2}h_{0}^{2}/\Delta f$ \\
  \begin{tabular}[t]{@{}l@{}}SNR relative to a face-on\\
       source with the same $h_0$\end{tabular}
    & unchanged
    & \begin{tabular}[t]{@{}l@{}}reduced by $\mu^{2}$,\\
       with $\mu^{2}=1$ only face-on\end{tabular} \\
\end{tabular}
\end{ruledtabular}
\end{minipage}
\end{center}
\twocolumngrid

\subsection{Orientation bias propagation}
\label{sec:orientation_chain}
The projection factor $\mu^2$ enters every physical quantity the
search reports. Three SNRs appear in what follows. They describe the
same source in the same data and differ only in the orientation at
which $\mu^2$ is evaluated.
\begin{itemize}
\item \textbf{Measured SNR}, $\hat{\rho}$. The SNR computed by the
  ASAF search, Eq.~\eqref{eq:snr}. It corresponds to the true
  orientation of the source, known for a hardware injection and
  unknown for a search candidate. The orientation suppression is
  therefore already contained in $\hat{\rho}$ and cannot be read off
  from it. Its null distribution calibrates the threshold
  $\rho_{\mathrm{thr}}$ of Eq.~\eqref{eq:threshold}.
\item \textbf{Recovered SNR}, $\rho_{\mathrm{rec}}(\iota,\psi)$. The
  SNR the same source would produce at an arbitrary orientation. It is
  a surface over the orientation plane, highest face-on, and
  $\hat{\rho}$ is its value at the true orientation.
\item \textbf{Reference SNR}, $\rho_{\mathrm{ref}}$. This is the peak
  of that surface, at $\mu^2 = 1$, that is, the SNR of a face-on
  source with the same $h_0$. It depends only on the intrinsic
  strength of the source and the detector noise, not on how the source
  is tilted. Two sources with the same $h_0$ therefore have the same
  $\rho_{\mathrm{ref}}$, whatever their orientation.
\end{itemize}

\paragraph{Strain.}
The intrinsic and effective strain amplitudes are related by
\begin{equation}
  h_0 = h_{\mathrm{eff}}/\sqrt{\mu^2}\,.
  \label{eq:h_chain}
\end{equation}
\paragraph{Signal-to-noise ratio.}
Because the radiometer estimates signal \emph{power}, the recovered
SNR is linear in $\mu^2$,
\begin{equation}
  \rho_{\mathrm{rec}}(\iota,\psi)
  = \mu^2(\iota,\psi)\,\rho_{\mathrm{ref}}\,.
  \label{eq:rho_chain}
\end{equation}
Since $\mu^2 \leq 1$, $\hat{\rho} \leq \rho_{\mathrm{ref}}$, with
equality only for a face-on source. Because published ASAF analyses
report only $\hat{\rho}$, a low value is ambiguous between an
intrinsically weak source and a strong one at an unfavorable
orientation. When the orientation is known, the projection factor
separates the two. For a PHWI the measured SNR is
$\hat{\rho} = \mu^2_{\mathrm{inj}}\rho_{\mathrm{ref}}$, which inverts
to $\rho_{\mathrm{ref}} = \hat{\rho}/\mu^2_{\mathrm{inj}}$. Since the
measured statistic has unit variance, the noise uncertainty on the
inferred reference SNR is
$\sigma(\rho_{\mathrm{ref}}) = 1/\mu^2_{\mathrm{inj}}$, and this is
the error bar quoted throughout.
\paragraph{Upper limits.}
An upper limit on the effective strain translates into a weaker limit
on the intrinsic amplitude,
\begin{equation}
  h_0^{\mathrm{UL}} = h_{\mathrm{eff}}^{\mathrm{UL}}/\sqrt{\mu^2}
  \geq h_{\mathrm{eff}}^{\mathrm{UL}}\,,
  \label{eq:ul_chain}
\end{equation}
with equality only for a face-on source.
\paragraph{Detection probability.}
With unit-variance noise on the recovered SNR, the probability of
exceeding the threshold $\rho_{\mathrm{thr}}$ is
\begin{equation}
  \Pdet(\iota,\psi)
  = \frac{1}{2}\,\operatorname{erfc}\!\biggl(
    \frac{\rho_{\mathrm{thr}} -
    \mu^2\rho_{\mathrm{ref}}}{\sqrt{2}}\biggr)\,.
  \label{eq:pdet_orientation}
\end{equation}
We summarize this by the detectability ratio
\begin{equation}
  \mathcal{R}(\iota,\psi)
  \equiv \frac{\mu^2(\iota,\psi)\,\rho_{\mathrm{ref}}}
         {\rho_{\mathrm{thr}}}\,.
  \label{eq:detectability_ratio}
\end{equation}
$\mathcal{R} \geq 1$ means that the mean recovered SNR reaches the
threshold, equivalently $\Pdet \geq 1/2$. We adopt this as the
detection criterion and call orientations with $\mathcal{R} < 1$
missed.

\paragraph{Orientation completeness.}
For a source of given $\rho_{\mathrm{ref}}$, the orientation
completeness is the fraction of isotropically distributed orientations
at which the source would be detected,
\begin{equation}
  \mathcal{C}
  = \frac{2}{\pi}\int_0^1 d\cos\iota
    \int_{-\pi/4}^{\pi/4} d\psi\;
    \Theta\bigl(\mathcal{R}(\iota,\psi) - 1\bigr)\,,
  \label{eq:completeness_def}
\end{equation}
where $\Theta$ is the step function. By the symmetry
$\iota \to \pi-\iota$ and the period $\pi/2$ of
Eq.~\eqref{eq:theorem} in $\psi$, this range covers all orientations.
We evaluate the integral on the centers of $N = 10 \times 10$ equal
cells in $(\cos\iota,\psi)$, so that $\mathcal{C} = \Ndet/N$, with
$\Ndet$ the number of cells with $\mathcal{R} \geq 1$.


\section{Orientation-marginalized upper limits}
\label{sec:marginalized_uls}
In the absence of a detection, the ASAF search sets Bayesian ULs on $h_{\mathrm{eff}}$. These are limits on $h_0$ only for a
circularly polarized source, $\mu^2 = 1$. For a source of unknown
orientation, the limit must be marginalized over orientation. We
follow the procedure of the published ASAF
analyses~\cite{KAGRA:2021rmt,LIGOScientific:2025bkz} and extend it
with the substitution $h_{\mathrm{eff}}^2 \to \mu^2(\iota,\psi)\,h_0^2$
of Sec.~\ref{sec:mu2}. The same marginalization has been applied in
directed radiometer
searches~\cite{Messenger:2010T1000195,Meyers:2018thesis,LIGOScientific:2025bkz}.

\subsection{Bayesian upper limits under the face-on assumption}
\label{sec:marg_posterior}
We adopt the hybrid frequentist-Bayesian approach of
Refs.~\cite{Matas:2020roi,KAGRA:2021rmt,LIGOScientific:2025bkz}. The
point estimate of Eq.~\eqref{eq:point_estimate} is treated as a
sufficient statistic, and Bayes' theorem gives the posterior of the
strain amplitude. We absorb the bin width into the estimators,
\begin{equation}
  \hat{\mathcal{P}} \equiv \hat{\mathcal{P}}_{\hat{n}}(f)\,\Delta f\,,
  \qquad
  \sigma_{\mathcal{P}} \equiv \sigma_{\hat{n}}(f)\,\Delta f\,,
  \label{eq:P_sigma_strain}
\end{equation}
so that $h_{\mathrm{eff}}^2 = \hat{\mathcal{P}}$ and
$\hat{\rho} = \hat{\mathcal{P}}/\sigma_{\mathcal{P}}$. The $95\%$
credible upper limit is defined by
\begin{equation}
  0.95
  = \int_0^{h^{\mathrm{UL}}_{\mathrm{eff}}}
    \mathrm{d}h_{\mathrm{eff}}\;
    \frac{\mathcal{L}(\hat{\mathcal{P}}\,|\,h_{\mathrm{eff}})\,
          p(h_{\mathrm{eff}})}
         {\int_0^{h_{\max}}
          \mathrm{d}h\;
          \mathcal{L}(\hat{\mathcal{P}}\,|\,h)\,p(h)}\,,
  \label{eq:ul_credible}
\end{equation}
where $\mathcal{L}$ is the likelihood and $p$ the prior.

The likelihood is Gaussian. The noise has variance
$\sigma_{\mathcal{P}}^2$. The amplitude calibration of the detectors
multiplies the signal power by an unknown factor
$\lambda$~\cite{Whelan:2012ur,LIGOScientific:2025bkz},
\begin{equation}
  \hat{\mathcal{P}} = \lambda\,h_{\mathrm{eff}}^2 + n\,,
  \quad
  n \sim \mathcal{N}(0,\sigma_{\mathcal{P}}^2)\,,
  \quad
  \lambda \sim \mathcal{N}(1,\sigma_\lambda^2)\,.
  \label{eq:measurement_model}
\end{equation}
Marginalizing over $\lambda$ adds $\sigma_\lambda^2 h_{\mathrm{eff}}^4$
to the variance,
\begin{equation}
  \mathcal{L}(\hat{\mathcal{P}}\,|\,h_{\mathrm{eff}})
  = \frac{1}{\sqrt{2\pi\bigl(\sigma_{\mathcal{P}}^2
      + \sigma_\lambda^2 h_{\mathrm{eff}}^4\bigr)}}\,
    \exp\!\biggl[-\frac{(\hat{\mathcal{P}} - h_{\mathrm{eff}}^2)^2}
      {2\bigl(\sigma_{\mathcal{P}}^2
      + \sigma_\lambda^2 h_{\mathrm{eff}}^4\bigr)}\biggr]\,.
  \label{eq:likelihood_pipeline}
\end{equation}
Here $\sigma_\lambda^2 = \sigma_{\mathrm{H}}^2 + \sigma_{\mathrm{L}}^2
+ \sigma_{\mathrm{H}}^2\sigma_{\mathrm{L}}^2$ combines the calibration
uncertainties of the two detectors. With $\sigma_{\mathrm{H}} = 0.026$
and $\sigma_{\mathrm{L}} = 0.0385$ for O2, and $0.0693$ and $0.041$ for
O4a~\cite{Cahillane:2017vkb,LIGOScientific:2025bkz}, we obtain
$\sigma_\lambda = 0.0465$ and $0.0806$. 

We adopt the uniform prior $h_{\mathrm{eff}} \sim \mathcal{U}[0,
h_{\max}]$ of the published
analyses~\cite{KAGRA:2021rmt,LIGOScientific:2025bkz}, with
\begin{equation}
  h_{\max} = 10\,\sqrt{\sigma_{\mathcal{P}}}\,.
  \label{eq:hmax_pipeline}
\end{equation}
The posterior is evaluated on $10^4$ uniform grid points over this
range. The UL is the value of $h_{\mathrm{eff}}$ at which the normalized
cumulative sum of the posterior reaches $0.95$. We use
the same implementation throughout, so our circular limits for the O4a
candidates reproduce the published ones~\cite{LIGOScientific:2025bkz}.

We measure $h$ in units of the noise scale,
$x = h/\sqrt{\sigma_{\mathcal{P}}}$. In these units $\sigma_{\mathcal{P}}$
drops out of the posterior, which then depends only on $\hat{\rho}$ and
$\sigma_\lambda$. The UL is therefore
\begin{equation}
  h^{\mathrm{UL}}_{\mathrm{eff}}
  = \sqrt{\sigma_{\mathcal{P}}}\;g(\hat{\rho},\sigma_\lambda)\,,
  \label{eq:ul_scaling}
\end{equation}
where $g$ is the UL on $x$. 

\subsection{Extension to a source of arbitrary orientation}
\label{sec:two_cases}
A CW source of amplitude $h_0$ at orientation $(\iota,\psi)$ produces
the cross-power $\mu^2 h_0^2$. The calibration factor multiplies this
power, and the noise is unchanged. Replacing $h_{\mathrm{eff}}^2$ by
$\mu^2 h_0^2$ in Eq.~\eqref{eq:likelihood_pipeline} gives
\begin{equation}
\begin{split}
  \mathcal{L}(\hat{\mathcal{P}}\,|\,h_0,\iota,\psi)
  = \frac{1}{\sqrt{2\pi\bigl(\sigma_{\mathcal{P}}^2
      + \sigma_\lambda^2 \mu^4 h_0^4\bigr)}}\\
  \times\exp\!\biggl[-\frac{(\hat{\mathcal{P}} - \mu^2 h_0^2)^2}
      {2\bigl(\sigma_{\mathcal{P}}^2
      + \sigma_\lambda^2 \mu^4 h_0^4\bigr)}\biggr]\,.
\end{split}
  \label{eq:likelihood}
\end{equation}
The orientation enters only through $\mu^2 h_0^2$. We consider three
prescriptions.

The \emph{circular prescription}, used by ASAF, sets $\mu^2 = 1$. It
gives the UL of Sec.~\ref{sec:marg_posterior}.

The \emph{at-target prescription} fixes $\mu^2$ at a known orientation.
The likelihood is then the circular one at
$h_{\mathrm{eff}} = \sqrt{\mu^2}\,h_0$. The posterior, its prior range,
and its $95\%$ quantile are all stretched by $1/\sqrt{\mu^2}$, so
\begin{equation}
  h_0^{\mathrm{UL}}(\iota,\psi)
  = \frac{h^{\mathrm{UL}}_{\mathrm{eff}}}{\sqrt{\mu^2(\iota,\psi)}}\,,
  \label{eq:ul_target}
\end{equation}
as in Eq.~\eqref{eq:ul_chain}. We use it for the hardware injections,
whose orientations are known, and to bound the UL of a candidate
over all orientations.

The \emph{marginalized prescription} treats the orientation as unknown.
The priors are uniform in $\cos\iota \in [0,1]$ and
$\psi \in [-\pi/4,\pi/4]$, which covers all orientations
(Sec.~\ref{sec:mu2}), and uniform in $h_0 \in
[0,h_{\max}^{\mathrm{gen}}]$. The posterior is
\begin{equation}
  p_{\mathrm{gen}}(h_0\,|\,\hat{\mathcal{P}})
  \propto \int_0^1\!\mathrm{d}(\cos\iota)
    \int_{-\pi/4}^{\pi/4}\!\frac{2\,\mathrm{d}\psi}{\pi}\;
    \mathcal{L}(\hat{\mathcal{P}}\,|\,h_0,\iota,\psi)\,,
  \label{eq:full_posterior}
\end{equation}
and the UL $h_0^{\mathrm{UL,gen}}$ follows from
Eq.~\eqref{eq:ul_credible}. Since $\mathcal{L}$ depends on the
orientation only through $\mu^2$, the double integral reduces to an
average over the prior distribution of $\mu^2$. We evaluate
Eq.~\eqref{eq:theorem} on a $1000 \times 1000$ grid in
$(\cos\iota,\psi)$ and histogram the values into $512$ bins with
centers $\mu^2_k$ and weights $w_k$, $\sum_k w_k = 1$. The posterior
becomes
\begin{equation}
  p_{\mathrm{gen}}(h_0\,|\,\hat{\mathcal{P}})
  \propto \sum_{k} w_k\,
    \mathcal{L}\bigl(\hat{\mathcal{P}}\,\big|\,h_0;\mu^2_k\bigr)\,,
  \label{eq:post_case2}
\end{equation}
a weighted sum of the likelihood of Eq.~\eqref{eq:likelihood} over the
orientation prior.

The prior range on $h_0$ requires care. An edge-on source must be
intrinsically louder than a face-on one to produce the same
cross-power. The posterior therefore extends to
$h_{\mathrm{eff}}/\sqrt{\mu^2_{\min}}$, about three times the face-on
scale. We take the range of the least favorable orientation,
\begin{equation}
  h_{\max}^{\mathrm{gen}}
  = \frac{10\,\sqrt{\sigma_{\mathcal{P}}}}{\sqrt{\mu^2_{\min}}}\,,
  \label{eq:hmax_gen}
\end{equation}
with $\mu^2_{\min}$ of Eq.~\eqref{eq:mumin}.

\subsection{Marginalized versus circular upper limits}
\label{sec:ul_cost}
The ratio $h_0^{\mathrm{UL,gen}}/h_0^{\mathrm{UL,circ}}$ measures the
cost of dropping the face-on assumption. With $\mu^2$ fixed at its mean
of $2/5$, it would be $\sqrt{5/2} \approx 1.58$. The actual ratio is
larger, because the $95\%$ limit is set by the upper tail of the
posterior. In Eq.~\eqref{eq:post_case2}, near-edge-on orientations
stretch the circular posterior the most, by $1/\sqrt{\mu^2_k}$, and
with a uniform prior on $h_0$ they also carry the largest weight,
$w_k/\sqrt{\mu^2_k}$.

The ratio is smallest for a pair consistent with noise, whose circular
posterior is already broad. It grows with $\hat{\rho}$ and with
$\varepsilon$, toward $1/\sqrt{\mu^2_5}$, where $\mu^2_5$ is the fifth
percentile of $\mu^2$ weighted by $w_k/\sqrt{\mu^2_k}$. The calibration
uncertainty keeps it slightly below this value. The targeted
narrowband radiometer of Ref.~\cite{LIGOScientific:2025bkz} reports a
ratio of about $2.5$, close to ours (Secs.~\ref{sec:ul_comparison}
and~\ref{sec:o4a_ul_ranges}). The closed form of
Eq.~\eqref{eq:theorem} explains this ratio and extends the
marginalization to every frequency-pixel pair of the ASAF search.
\section{Application to O2 hardware injections}
\label{sec:orientation_results}
We now apply the formalism to four PHWIs injected during O2, with
known source parameters~\cite{VIRGO:2009quh,Biwer:2016oyg}.
They lie at four sky positions, and their polarization ranges from
nearly circular (face-on) to nearly linear (edge-on). We order them by
decreasing $|\cos\iota|$.

\subsection{Injection properties and measured quantities}
\label{sec:test_cases}
All results use $\nside = 16$, $\rho_{\mathrm{thr}} = 6.19$, and
$\sigma_\lambda = 0.0465$, the combined O2 amplitude calibration
uncertainty of Sec.~\ref{sec:marg_posterior}. The injected
orientations are known exactly from the hardware-injection parameter
files~\cite{Biwer:2016oyg}. The only systematic in $\mu^2$ is
therefore the use of the pixel center instead of the true source
position. Its effect on $\mu^2$ is negligible compared with the noise.

Table~\ref{tab:faceon_measurements} collects the results. The four
injections span the full range of polarization. Injection~1 is nearly
face-on ($|\cos\iota| = 0.988$) and circularly polarized. Injection~4
is nearly edge-on ($|\cos\iota| = 0.081$) and linearly polarized.
Injections~2 and~3 lie in between. The measured SNR follows the same
order. It falls from $\hat{\rho} = 26.63$ for Injection~1 to
$\hat{\rho} = 5.20$ for Injection~4, which is below threshold. The
projection factor explains this difference: $\mu^2 = 0.976$ for
Injection~1, but only $0.167$ for Injection~4. Dividing $\hat{\rho}$
by $\mu^2$ gives the face-on-equivalent SNRs
$\rho_{\mathrm{ref}} = 27.3 \pm 1.0$, $45.2 \pm 2.3$, $56.8 \pm 3.4$,
and $31.1 \pm 6.0$ for Injections~1 to~4. Since $\hat{\rho}$ has unit
variance, $\sigma(\rho_{\mathrm{ref}}) = 1/\mu^2$
(Sec.~\ref{sec:orientation_chain}).

The injected amplitudes are also known, so $\mu^2$ can be tested
directly. Dividing $h_{\mathrm{eff}}$ by $\sqrt{\mu^2}$ gives the
recovered $h_0$ of Table~\ref{tab:faceon_measurements}, with
uncertainty from $\sigma(h_{\mathrm{eff}}) =
h_{\mathrm{eff}}/(2\hat{\rho})$. For Injections~1 and~3, the recovered
$h_0$ agrees with the injected value. For Injections~2 and~4, it is
lower by factors of $2.2$ and $1.3$. This deficit is not caused by the
orientation. It comes from two known losses of cross-power in the
analyzed bin:
\begin{itemize}
  \item \emph{Doppler loss.} The orbital motion of the Earth spreads
  the line over neighboring bins. The line of Injection~4, at
  $108.857$~Hz, lies only $2$~mHz below the upper edge of its
  $1/32$~Hz bin, so about $40\%$ of its power moves into the next bin.
  Injection~2, at $763.85$~Hz, loses about half of its power in the
  same way.
  \item \emph{Pixel-phase loss.} The radiometer phase is evaluated at
  the pixel center, $2.1^{\circ}$ from Injection~2. This costs it
  about half of its remaining power.
\end{itemize}
Correcting for these losses gives $h_0 \approx 2.9\times10^{-24}$ for
Injection~2 and $8.2\times10^{-25}$ for Injection~4, close to the
injected values. The losses lower $\hat{\rho}$, $h_{\mathrm{eff}}$, and
the quantities derived from them. They do not change $\mu^2$, or the
ratios between the UL prescriptions of
Sec.~\ref{sec:ul_comparison}.

\begin{table*}[!t]
\squeezetable
\caption{How source orientation affects what the ASAF
    formalism measures. The top block gives the known source
    parameters, with $\psi$ in the convention of
    Eqs.~\eqref{eq:Fplus} and~\eqref{eq:Fcross}. The middle block shows the uncorrected
    ASAF output, with $\sigma(h_{\mathrm{eff}}) =
    h_{\mathrm{eff}}/(2\hat{\rho})$. The bottom block applies the
    orientation correction, with $\mu^2 = h_{\mathrm{eff}}^2/h_0^2$.
    Injection~4 falls below threshold. The recovered $h_0$ of Injections~2 and~4 is low because of the Doppler
and pixel-phase losses of Sec.~\ref{sec:test_cases}.}
\label{tab:faceon_measurements}
\begin{ruledtabular}
\renewcommand{\arraystretch}{1.2}
\begin{tabular}{lcccc}
    & Inj.\,1 & Inj.\,2 & Inj.\,3 & Inj.\,4 \\
    \colrule
    \multicolumn{5}{c}{\textit{How is the source oriented?}} \\
    \colrule
     Frequency bin $f$ [Hz]
      & 26.34 & 763.84 & 52.81 & 108.84 \\
    Sky position $(\mathrm{RA},\mathrm{Dec})$
      & $(14.8^{\mathrm{h}},\,{+}42.9^{\circ})$
      & $(13.3^{\mathrm{h}},\,{+}75.7^{\circ})$
      & $(20.2^{\mathrm{h}},\,{-}83.8^{\circ})$
      & $(11.9^{\mathrm{h}},\,{-}33.4^{\circ})$ \\
    Inclination $\iota_{\mathrm{inj}}$ [deg]
      & $171.0$ & $128.3$ & $62.4$ & $94.6$ \\
    Polarization angle $\psi_{\mathrm{inj}}$ [deg]
      & ${-}35.2$ & ${+}0.5$ & ${+}20.9$ & ${-}25.5$ \\
    $\cos\iota_{\mathrm{inj}}$
      & $-0.988$ & $-0.619$ & ${+}0.463$ & $-0.081$ \\
    Polarization state
      & circular & elliptical & elliptical & linear \\
    $|A_{\times}/A_{+}|$
      & 1.00 & 0.90 & 0.76 & 0.16 \\
    Injected strain $h_0^{\mathrm{inj}}$
      & $4.00 \times 10^{-24}$
      & $3.01 \times 10^{-24}$
      & $1.46 \times 10^{-24}$
      & $8.23 \times 10^{-25}$ \\
    \colrule
    \multicolumn{5}{c}{\textit{What does the ASAF measure without
      correction?}} \\
    \colrule
    $h_{\mathrm{eff}}$
      & $(3.91 \pm 0.07) \times 10^{-24}$
      & $(8.83 \pm 0.23) \times 10^{-25}$
      & $(7.99 \pm 0.24) \times 10^{-25}$
      & $(2.57 \pm 0.25) \times 10^{-25}$ \\
    Measured SNR $\hat{\rho}$
      & 26.63 & 19.46 & 16.57 & 5.20 \\
    Exceeds $\rho_{\mathrm{thr}} = 6.19$?
      & yes & yes & yes & no \\
    \colrule
    \multicolumn{5}{c}{\textit{What does the corrected analysis
      reveal?}} \\
    \colrule
    $\mu^2(\iota_{\mathrm{inj}},\psi_{\mathrm{inj}})$
      & 0.976 & 0.431 & 0.292 & 0.167 \\
    Correction factor $1/\!\sqrt{\mu^2}$
      & 1.01 & 1.52 & 1.85 & 2.45 \\
    Face-on SNR $\rho_{\mathrm{ref}} = \hat{\rho}/\mu^2$
      & $27.3 \pm 1.0$
      & $45.2 \pm 2.3$
      & $56.8 \pm 3.4$
      & $31.1 \pm 6.0$ \\
    Recovered $h_0 = h_{\mathrm{eff}}/\!\sqrt{\mu^2}$
      & $(3.96 \pm 0.07) \times 10^{-24}$
      & $(1.34 \pm 0.03) \times 10^{-24}$
      & $(1.48 \pm 0.04) \times 10^{-24}$
      & $(6.28 \pm 0.60) \times 10^{-25}$ \\
\end{tabular}
\end{ruledtabular}
\end{table*}

\subsection{Projection-factor maps}
\label{sec:mu2_maps}
\begin{figure}[!t]
  \centering
  \includegraphics[width=\columnwidth]{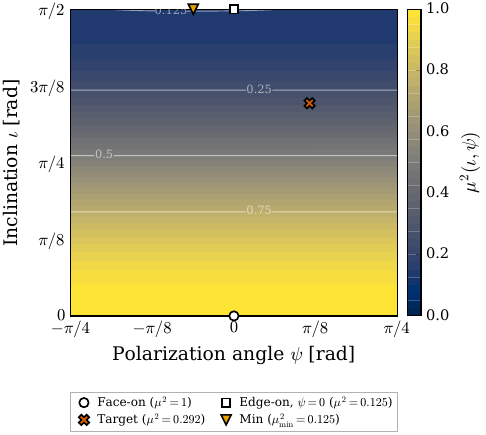}\\[2pt]
  \includegraphics[width=\columnwidth]{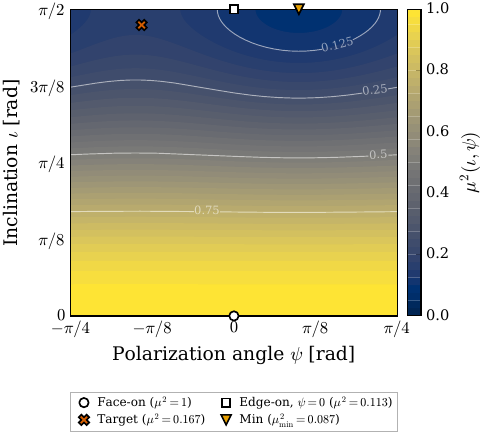}
  \caption{Projection factor $\mu^2(\iota,\psi)$ for Injection~3
    (top, Dec~$= -83.84^{\circ}$, near the south celestial pole)
    and Injection~4 (bottom, Dec~$= -33.44^{\circ}$), on a common
    color scale. Markers show the face-on (circle), target
    (cross), and edge-on (square) orientations, and the minimum
    $\mu^2_{\min} = (1-\varepsilon)/8$ at $\psi^*$ (triangle).
    Injection~3 has $\varepsilon \approx 10^{-3}$, and its
    contours barely depend on $\psi$. Injection~4 has
    $\varepsilon = 0.302$, and its contours vary with $\psi$. Both
    average to $\langle\mu^2\rangle = 2/5$.}
  \label{fig:mu2_contour}
\end{figure}

Figure~\ref{fig:mu2_contour} shows $\mu^2(\iota,\psi)$ across the
full orientation plane for Injections~3 and~4. The two panels differ in one
aspect, the presence or absence of structure along the $\psi$
axis. Equation~\eqref{eq:theorem} explains this difference. Its first term
depends only on $\iota$. The $\psi$ structure comes from the second
term, whose depth $\varepsilon$ depends on the sky position, and with
uniform weights only on the declination
(Appendix~\ref{app:closedform}). A map shows $\psi$ structure only
where the coefficients $C^{(1)}_t$ and $C^{(2)}_t$ do not cancel in
the daily sums of Eq.~\eqref{eq:timesums}, so that $\varepsilon$ is
not small.

Injection~3 (top panel, $\mathrm{RA} = 20.2^{\mathrm{h}}$,
Dec~$= -83.84^{\circ}$) lies $6^{\circ}$ from the south celestial
pole. For a source at the pole, the rotation of the Earth turns both
detectors about the line of sight. Over a sidereal day this sweeps the
polarization angle seen by the detectors uniformly, and the modulation
coefficients cancel in the sum. Here $\varepsilon \approx 10^{-3}$.
The modulation term of Eq.~\eqref{eq:theorem} then averages out, the
contours become nearly horizontal bands that depend only on $\iota$,
and the minimum $\mu^2_{\min} = 0.125 = 1/8$ is the analytic edge-on
limit of Eq.~\eqref{eq:mu2_psi_avg}.

Injection~4 (bottom panel, $\mathrm{RA} = 11.9^{\mathrm{h}}$,
Dec~$= -33.44^{\circ}$) lies at mid-declination. There the
rotation of the Earth changes the projected detector geometry as well
as the angle, and the coefficients cancel only in part. The
modulation depth is $\varepsilon = 0.302$, and the contours vary with
$\psi$ with period $\pi/2$. The minimum is
$\mu^2_{\min} = (1-\varepsilon)/8 = 0.087$, from Eq.~\eqref{eq:mumin}.
It lies below the $\psi$-averaged edge-on value $1/8$. The
coefficients $\varepsilon$ and $\psi^*$ of all four injections are
listed in Table~\ref{tab:completeness}.

\subsection{Detection probability and orientation completeness}
\label{sec:detection_results}
For each injection we evaluate $\rho_{\mathrm{rec}}$ and $\Pdet$ of
Eqs.~\eqref{eq:rho_chain} and~\eqref{eq:pdet_orientation} at four
reference orientations: face-on ($\mu^2 = 1$), the true injection
orientation (target), edge-on ($\iota = \pi/2$, $\psi = 0$), and the
isotropic average ($\mu^2 = 2/5$). Face-on and at the isotropic
average, all four injections have $\Pdet = 1.000$. Even with $40\%$ of
the face-on power, every source stays well above threshold.
Table~\ref{tab:completeness} lists the target and edge-on values.
Edge-on, only Injection~3 stays above threshold, with
$\rho_{\mathrm{rec}} = 7.10$ and $\Pdet = 0.820$. Even Injection~1
($\rho_{\mathrm{ref}} = 27.3$) drops to $\Pdet = 0.001$. At its true
orientation, Injection~4 has $\Pdet = 0.161$.

\begin{figure*}[!t]
  \centering
  \includegraphics[width=\textwidth]{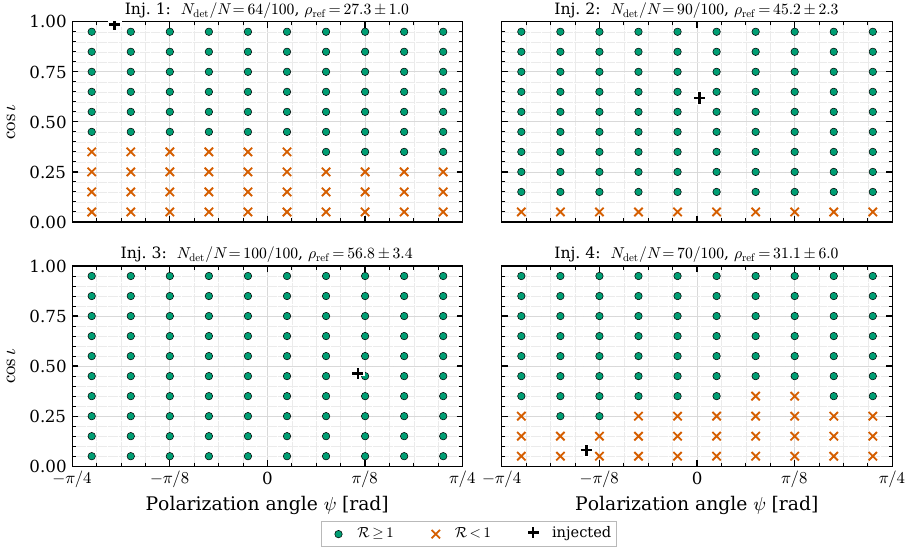}
  \caption{Detectability grid on the $(\psi, \cos\iota)$ plane for the
    four injections, evaluated at $\rho_{\mathrm{thr}} = 6.19$. Filled
    circles mark orientations where $\mathcal{R} \geq 1$ (detected).
    Open crosses mark $\mathcal{R} < 1$ (missed). The black cross
    marks the true injection orientation. Panel titles quote
    $\Ndet/N$ and
    $\rho_{\mathrm{ref}} \pm \sigma(\rho_{\mathrm{ref}})$.}
  \label{fig:detectability_grid}
\end{figure*}

\begin{table}[!h]
  \squeezetable
  \caption{Detection statistics for the four O2 injections at
    $\rho_{\mathrm{thr}} = 6.19$. The coefficients $\varepsilon$ and
    $\psi^*$ are those of Eq.~\eqref{eq:theorem} at each sky
    position, $\mu^2_{\min} = (1-\varepsilon)/8$ is the pointwise
    minimum of Eq.~\eqref{eq:mumin}, and
    $\rho_{\mathrm{ref,crit}} = \rho_{\mathrm{thr}}/\mu^2_{\min}$.
    Edge-on rows refer to $(\iota,\psi) = (90^{\circ},0^{\circ})$.
    The completeness $\mathcal{C} = \Ndet/N$ counts the cells of the
    $10 \times 10$ grid in $(\cos\iota,\psi)$ with
    $\mathcal{R} \geq 1$, and $\iota_{\max}^{\exists\psi}$ solves
    Eq.~\eqref{eq:imaxexists}.}
  \label{tab:completeness}
  \begin{ruledtabular}
  \begin{tabular}{lcccc}
    & Inj.\,1 & Inj.\,2 & Inj.\,3 & Inj.\,4 \\
    \colrule
    $\rho_{\mathrm{ref}}$
      & $27.3 \pm 1.0$
      & $45.2 \pm 2.3$
      & $56.8 \pm 3.4$
      & $31.1 \pm 6.0$ \\[2pt]
    $\varepsilon$
      & 0.202 & 0.004 & 0.001 & 0.302 \\[2pt]
    $\psi^*$ [deg]
      & $-18.0$ & $-18.4$ & $-11.2$ & $+17.9$ \\[2pt]
    $\mu^2_{\min}$
      & 0.100 & 0.125 & 0.125 & 0.087 \\[2pt]
    $\rho_{\mathrm{ref,crit}}$
      & 62.0 & 49.7 & 49.5 & 70.9 \\[2pt]
    \colrule
    Edge-on $\rho_{\mathrm{rec}}$
      & 3.20 & 5.64 & 7.10 & 3.52 \\[2pt]
    $\Pdet$ at target
      & 1.000 & 1.000 & 1.000 & 0.161 \\[2pt]
    Edge-on $\Pdet$
      & 0.001 & 0.292 & 0.820 & 0.004 \\[2pt]
    \colrule
    $\mathcal{C} = \Ndet/N$
      & 64/100 & 90/100 & 100/100 & 70/100 \\[2pt]
    $\iota_{\max}^{\exists\psi}$
      & $70.9^{\circ}$ & $82.9^{\circ}$
      & $90.0^{\circ}$ & $76.7^{\circ}$ \\[2pt]
    $\mathcal{R}$ at target
      & 4.30 & 3.14 & 2.68 & 0.84 \\
  \end{tabular}
  \end{ruledtabular}
\end{table}
We evaluate $\mathcal{R}$ on the $10 \times 10$ grid in
$(\cos\iota, \psi)$ of Eq.~\eqref{eq:completeness_def}.
Figure~\ref{fig:detectability_grid} shows the grid for all four
injections and Table~\ref{tab:completeness} summarizes it. Only
Injection~3 is detected at every orientation, since its
$\rho_{\mathrm{ref}} = 56.8$ exceeds $\rho_{\mathrm{ref,crit}} = 49.5$.
The other three are missed at the least favorable orientations.
Injection~4 is missed even at its true orientation
($\mathcal{R} = 0.84$), partly because of the bin-edge loss of
Sec.~\ref{sec:test_cases}. Without that loss, its measured SNR would
be about $9$, above threshold.

Injection~4 also shows how much the interpretation depends on the
orientation. Read at face value, $\hat{\rho} = 5.20$ describes a
source that is almost never detected ($\langle\Pdet\rangle = 0.01$).
Divided by $\mu^2$, it describes a source six times stronger, with
$\rho_{\mathrm{ref}} = 31.1$, detected at $70\%$ of orientations
($\langle\Pdet\rangle = 0.71$). This correction needs the orientation,
which is known here only because the signals are injections. For a search candidate the orientation is unknown, so $\mu^2$ cannot
be divided out. Its intrinsic strength can then only be bounded, by
an UL on $h_0$ marginalized over orientation
(Sec.~\ref{sec:ul_comparison}).

\subsection{Upper limits}
\label{sec:ul_comparison}
Table~\ref{tab:ul_values} gives the $95\%$ credible ULs on
$h_0$ under the three prescriptions of Sec.~\ref{sec:two_cases}, with
$\sigma_\lambda = 0.0465$. The marginalized ULs are $2.69$ to $2.85$
times the circular ones, well above $\sqrt{5/2} \approx 1.58$
(Sec.~\ref{sec:ul_cost}).
\begin{table}[!h]
  \squeezetable
  \caption{$95\%$ credible ULs on $h_0$ for the four O2
    hardware injections in units of
    $10^{-24}$ (Injection~1) and $10^{-25}$ (Injections~2 to~4), under
    the three prescriptions of Sec.~\ref{sec:two_cases}.}
  \label{tab:ul_values}
  \begin{ruledtabular}
  \begin{tabular}{lcccc}
    & Inj.\,1 & Inj.\,2 & Inj.\,3 & Inj.\,4 \\
    & $[10^{-24}]$ & $[10^{-25}]$ & $[10^{-25}]$ & $[10^{-25}]$ \\
    \colrule
    \multicolumn{5}{c}{\textit{$h_0^{\mathrm{UL}}$ at $95\%$
      credible level}} \\
    \colrule
    Circular ($\mu^2 = 1$)
      & 4.11 & 9.34 & 8.49 & 2.94 \\[2pt]
    At target
      & 4.16 & 14.2 & 15.7 & 7.20 \\[2pt]
    Marginalized
      & 11.7 & 25.2 & 22.8 & 8.16 \\[2pt]
    Grid minimum
      & 4.33 & 9.83 & 8.93 & 3.10 \\[2pt]
    Grid maximum
      & 12.8 & 26.3 & 23.8 & 9.76 \\[2pt]
    \colrule
    \multicolumn{5}{c}{\textit{Ratio to the circular upper
      limit}} \\
    \colrule
    Marginalized\,/\,Circular
      & 2.85 & 2.70 & 2.69 & 2.77 \\[2pt]
    Target\,/\,Circular
      & 1.01 & 1.52 & 1.85 & 2.45 \\
  \end{tabular}
  \end{ruledtabular}
\end{table}

\begin{figure}[!h]
  \centering
  \includegraphics[width=0.92\columnwidth]{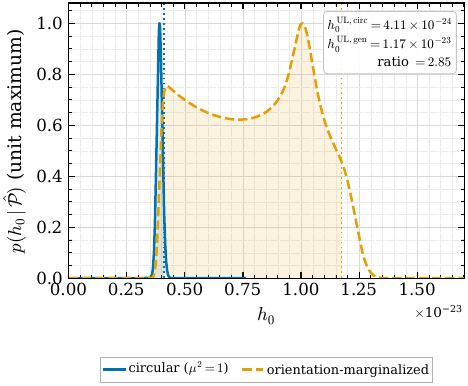}
  \caption{Posterior on $h_0$ for Injection~1, circular (solid) and
    marginalized over orientation (dashed), each normalized to unit
    maximum. Vertical lines mark the $95\%$ quantiles, which differ by
    a factor of $2.85$.}
  \label{fig:ul_posteriors}
\end{figure}

Figure~\ref{fig:ul_posteriors} shows the two posteriors for
Injection~1, which has the highest SNR. The marginalized posterior
extends to $h_{\mathrm{eff}}/\sqrt{\mu^2_{\min}} \simeq
3\,h_{\mathrm{eff}}$, and its ratio of $2.85$. As a clear
detection, Injection~1 is better described by an interval. At the
injected orientation, the $90\%$ credible interval
$[3.77, 4.16]\times10^{-24}$ contains the injected
$4.00\times10^{-24}$. We quote $95\%$ quantiles because the search
reports them at every frequency-pixel pair. The marginalized posterior
agrees in shape with Ref.~\cite{Meyers:2018thesis}.

The target-to-circular ratio is $1/\sqrt{\mu^2_{\mathrm{target}}}$,
from $1.01$ for the nearly face-on Injection~1 to $2.45$ for the nearly
edge-on Injection~4. The at-target limit of Injection~4 lies below the injected $h_0$
because it covers only the power in the analyzed bin, about $60\%$ of
the signal (Sec.~\ref{sec:test_cases}). Injection~2 lies above $160$~Hz, where
ASAF quotes no ULs~\cite{LIGOScientific:2025bkz}. We keep it
because its losses cancel in the ratios. Matched-filter CW searches
reach deeper ULs by tracking the Doppler modulation, at much higher computational
cost~\cite{Wette:2023dom,Riles:2022wwz}.


\section{Application to O4a follow-up candidates}
\label{sec:o4a_application}

We now apply the formalism to the sub-threshold candidates of the O4a
ASAF search~\cite{LIGOScientific:2025bkz}, whose orientation is
unknown. We ask two questions: what range of $h_0$ is compatible with
the measured cross-power (Sec.~\ref{sec:o4a_ul_ranges}), and, if a
candidate is a real source, how inclined it could be and still be
detected (Sec.~\ref{sec:o4a_iota_max}). Section~\ref{sec:o4a_injection}
tests both on a PHWI recovered in O4a, whose orientation is known.
Neither question affects the significance of the candidates, whose
threshold and $p$-values remain as published.

\subsection{O4a data set}
\label{sec:o4a_data}
The O4a ASAF analysis~\cite{LIGOScientific:2025bkz} used H1-L1 data,
whose quality and sensitivity in O4a are described in
Refs.~\cite{Capote:2024rmo,LIGO:2024kkz}. It selected $505$
sub-threshold candidates for follow-up, between $23.8$ and
$1724.5$~Hz, with $3.59 \leq \hat{\rho} \leq 5.47$. For each candidate,
the released catalog gives the sky position, $h_{\mathrm{eff}}$ with its
uncertainty, and the published circular $95\%$ UL. The PHWI at
$52.8125$~Hz is not in this list, and we treat it separately in
Sec.~\ref{sec:o4a_injection}.

\subsection{UL ranges for the candidates}
\label{sec:o4a_ul_ranges}
For each candidate we first reproduce the published circular UL, with
the prescription of Sec.~\ref{sec:marg_posterior} and
$\sigma_\lambda = 0.0806$. We then compute two more limits. The first
is the marginalized UL of Eq.~\eqref{eq:post_case2}, which assumes an
unknown, isotropically distributed orientation. The second is the
worst-orientation UL, which assumes the least favorable orientation.
This is the edge-on case at $\psi^*$, and by Eq.~\eqref{eq:ul_target}
it equals the circular UL times $\sqrt{8/(1-\varepsilon)}$.

Table~\ref{tab:o4a_ul_ranges} lists five representative candidates.
Over all $505$ candidates, the marginalized UL is $2.54$ to $2.87$
times the circular one, with a median of $2.68$. This is the range
expected from Sec.~\ref{sec:ul_cost}. The worst-orientation UL is
$2.83$ times the circular one near the celestial poles. It rises to
$3.63$ times where $\varepsilon = 0.39$. The published circular UL is
therefore the lowest value of a range. The highest value is about
three times larger. We propose that ASAF analyses in future observing
runs report this range and the marginalized UL for each candidate, instead of the circular UL alone.

\begin{table}[!h]
  \squeezetable
  \caption{$95\%$ credible ULs on $h_0$ for five O4a candidates. The columns give the circular UL, the UL marginalized over an
    isotropic orientation, and the UL at the least favorable
    orientation. Ratios to the circular UL are in parentheses.}
  \label{tab:o4a_ul_ranges}
  \begin{ruledtabular}
  \begin{tabular}{lccccc}
        $f$ [Hz] & $\hat{\rho}$ & $\varepsilon$
    & Circular & Marginalized & Worst \\
    & & & $[10^{-25}]$ & $[10^{-25}]$ & $[10^{-25}]$ \\
    \colrule
    23.78125  & 3.59 & 0.181 & 10.4 & 26.8 (2.58) & 32.4 (3.13) \\[2pt]
    38.68750  & 3.80 & 0.001 & 3.54 & 8.99 (2.54) & 10.0 (2.83) \\[2pt]
    92.81250  & 5.22 & 0.122 & 1.68 & 4.39 (2.61) & 5.08 (3.02) \\[2pt]
    228.96875 & 4.43 & 0.394 & 1.32 & 3.76 (2.84) & 4.81 (3.63) \\[2pt]
    615.71875 & 5.47 & 0.005 & 1.57 & 4.06 (2.58) & 4.46 (2.84) \\
  \end{tabular}
  \end{ruledtabular}
\end{table}

\subsection{Maximum detectable inclination}
\label{sec:o4a_iota_max}
If a candidate is a real source, a source of the same strength but
more inclined produces less power. Beyond some inclination
$\iota_{\max}$, its SNR falls
below $\rho_{\mathrm{thr}}$. Computing $\iota_{\max}$ requires $\rho_{\mathrm{ref}}$, but only
$\hat{\rho} = \mu^2\rho_{\mathrm{ref}}$ is measured, and $\mu^2$ is
unknown. We therefore assume that the candidate is a source at a
typical orientation, $\mu^2 = \langle\mu^2\rangle$, so that
\begin{equation}
  \rho_{\mathrm{ref}}
    = \hat{\rho}\,/\,\langle\mu^2\rangle
    = 2.5\;\hat{\rho}\,.
  \label{eq:o4a_rho_corr}
\end{equation}

This value describes a hypothetical source. It is not a detection
claim, and it is not compared with $\rho_{\mathrm{thr}}$, which applies
to the measured $\hat{\rho}$. We now tilt this source away from
face-on. By Eq.~\eqref{eq:rho_chain}, its recovered SNR is
$\mu^2\rho_{\mathrm{ref}}$. With the $\psi$-averaged $\mu^2$ of
Eq.~\eqref{eq:mu2_psi_avg}, the SNR falls to the threshold at the
inclination $\iota_{\max}$ given by
\begin{equation}
  \frac{1 + 6\cos^2\!\iota_{\max} + \cos^4\!\iota_{\max}}{8}
  = \frac{\rho_{\mathrm{thr}}}{\rho_{\mathrm{ref}}}\,,
  \label{eq:iota_max_condition}
\end{equation}
which gives
\begin{equation}
  \cos^2\!\iota_{\max}
  = -3 + \sqrt{8 + 8\,\rho_{\mathrm{thr}}/\rho_{\mathrm{ref}}}\,.
  \label{eq:cosi_max_solution}
\end{equation}
The exact $\iota_{\max}$ also depends on $\psi$. Two limiting cases
bound it. Setting $\varepsilon = 1$ in Eq.~\eqref{eq:theorem} gives
$\mu^2 = \mathcal{A}_+^2$ at the best $\psi$ and
$\mu^2 = \mathcal{A}_\times^2$ at the worst. Real sky positions have
$\varepsilon \leq 0.395$, so the true $\iota_{\max}$ lies between these
two envelopes.

\begin{figure}[!h]
  \centering
  \includegraphics[width=\columnwidth]{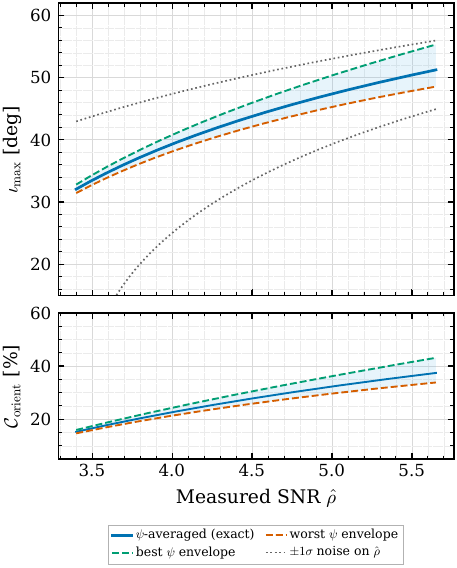}
  \caption{Maximum detectable inclination $\iota_{\max}$ (upper
    panel) and orientation coverage
    $\mathcal{C}_{\mathrm{orient}} = 1 - \cos\iota_{\max}$ (lower
    panel) as a function of $\hat{\rho}$ over the range of the
    O4a candidates. The shaded band lies between the best-$\psi$ and
    worst-$\psi$ envelopes (dashed curves). Dotted curves show the
    effect of $\pm 1\sigma$ noise on $\hat{\rho}$.}
  \label{fig:o4a_iota_max}
\end{figure}

Table~\ref{tab:o4a_iota_max} and Fig.~\ref{fig:o4a_iota_max} show the
results. The $\psi$-averaged $\iota_{\max}$ ranges from $34.7^{\circ}$
to $50.3^{\circ}$, with a median of $43.9^{\circ}$. No candidate
reaches $60^{\circ}$, even at the best $\psi$. For isotropic
orientations $\cos\iota$ is uniform, so the fraction of sources with
$\iota < \iota_{\max}$ is the orientation coverage
$\mathcal{C}_{\mathrm{orient}} = 1 - \cos\iota_{\max}$. It is only
$18\%$ to $36\%$. A source as strong as these candidates is therefore
detected only if it lies within about $35^{\circ}$ to $50^{\circ}$ of
face-on.
\begin{table}[!h]
  \caption{Maximum detectable inclination $\iota_{\max}$ for the
    $504$ O4a follow-up candidates, for a source with
    $\rho_{\mathrm{ref}} = 2.5\,\hat{\rho}$. Best $\psi$ and worst
    $\psi$ are the two envelopes. The last row counts the
    candidates with $\iota_{\max} \geq 45^{\circ}$.}
  \label{tab:o4a_iota_max}
  \squeezetable
  \begin{ruledtabular}
  \begin{tabular}{lccc}
    Statistic
    & $\iota_{\max}$ ($\psi$-avg)
    & $\iota_{\max}$ (best $\psi$)
    & $\iota_{\max}$ (worst $\psi$) \\
    \colrule
    Minimum    & $34.7^{\circ}$ & $35.6^{\circ}$ & $33.9^{\circ}$ \\
    Median     & $43.9^{\circ}$ & $46.1^{\circ}$ & $42.2^{\circ}$ \\
    Mean       & $44.0^{\circ}$ & $46.3^{\circ}$ & $42.3^{\circ}$ \\
    Maximum    & $50.3^{\circ}$ & $54.0^{\circ}$ & $47.7^{\circ}$ \\
    \colrule
    $N(\iota_{\max} \geq 45^{\circ})$
      & $128/504$ & $397/504$ & $30/504$ \\
  \end{tabular}
  \end{ruledtabular}
\end{table}

\subsection{The PHWI at 52.8125 Hz}
\label{sec:o4a_injection}

Of the O4a PHWIs~\cite{Baxi:2026wbr}, only the one at $52.8125$~Hz was
recovered as a follow-up candidate. Its parameters are
$\mathrm{Dec} = -83.8^{\circ}$, $\iota = 62.4^{\circ}$,
$\psi = +20.9^{\circ}$, and $h_0 = 3.99\times10^{-25}$. The O4a ASAF
analysis recovered it at $\hat{\rho} = 4.25$, below the detection
threshold~\cite{LIGOScientific:2025bkz}. Since the injection lies close
to the celestial pole, $\varepsilon \approx 10^{-3}$ and $\mu^2$
depends on the orientation alone, giving $\mu^2 = 0.292$.

Viewed face-on, the same source would have
$\rho_{\mathrm{ref}} = 14.6 \pm 3.4$, well above threshold. For this
strength $\iota_{\max} = 52.2^{\circ}$, and the injected inclination
of $62.4^{\circ}$ lies beyond it. This is why the injection was missed. The recovered amplitude agrees with this formalism. A face-on reading
would give $h_{\mathrm{eff}} = h_0$, whereas Eq.~\eqref{eq:h_chain}
predicts $\sqrt{\mu^2}\,h_0 = 2.15\times10^{-25}$. At the known
orientation, the UL is $1.85$ times the circular one. The
marginalized and worst-orientation limits are $2.56$ and $2.83$ times
the circular one.

\section{Conclusions}
\label{sec:conclusions}
The ASAF radiometer gives equal weight to the $+$ and $\times$
polarization states. This is optimal for an unpolarized SGWB, but it
underestimates the strength of CW sources. The projection factor
$\mu^2(\iota,\psi) = h_{\mathrm{eff}}^2/h_0^2$~\cite{Messenger:2015tga,Meyers:2018thesis}
measures this effect. It is the ratio of the cross-power the
radiometer measures to that of a face-on source with the same $h_0$.
We showed that its isotropic average is exactly $2/5$, independent of
sky position, detector geometry, noise weighting, and frequency, and
that its dependence on $\psi$ has the closed form of
Eq.~\eqref{eq:theorem}.

The four O2 hardware injections confirm the formalism. The measured
SNR is only a fraction $\mu^2$ of the face-on value. For Injection~4,
$\hat{\rho} = 5.20$ falls below threshold, although
$\rho_{\mathrm{ref}} = 31.1 \pm 6.0$. Dividing $h_{\mathrm{eff}}$ by
$\sqrt{\mu^2}$ recovers the injected $h_0$, once the known Doppler and
pixel-phase losses are included. ULs marginalized over orientation, as
in Refs.~\cite{Messenger:2010T1000195,Meyers:2018thesis}, are $2.7$ to
$2.9$ times the circular ones for the injections. They follow from the
published ASAF prescription with the single substitution
$h_{\mathrm{eff}}^2 \to \mu^2 h_0^2$.

For the $505$ O4a follow-up candidates, the correction leaves every
$p$-value unchanged but changes their interpretation. Each published
circular UL is the lowest value of a range over orientation whose
highest value is about three times larger, and the marginalized UL is
$2.5$ to $2.9$ times the circular one. A source as strong as these
candidates is detected only within $35^{\circ}$ to $50^{\circ}$ of
face-on, which covers only $18\%$ to $36\%$ of isotropic orientations.
The PHWI at $52.8125$~Hz illustrates this. It was recovered at
$\hat{\rho} = 4.25$, although viewed face-on it would have had
$\rho_{\mathrm{ref}} = 14.6 \pm 3.4$, a confident detection. Its
inclination of $62.4^{\circ}$ lies beyond the detectable
$52.2^{\circ}$, and its recovered amplitude follows the orientation
prediction, not the face-on one.

These results have three practical implications. First, face-on ULs
are $2$ to $3$ times lower than those for a randomly oriented source.
We recommend quoting them as a range over orientation. Second,
$\iota_{\max}$ and $\mathcal{C}_{\mathrm{orient}}$ should be reported
alongside $\hat{\rho}$ and $h_{\mathrm{eff}}$. Third, a low
$\hat{\rho}$ does not imply a weak source. The significance of a
candidate is unchanged, but the intrinsic strength behind a given
$\hat{\rho}$ can be up to $1/\mu^2_{\min} \approx 8$ to $13$ times
larger. Selecting candidates by $\hat{\rho}$ alone therefore favors
face-on sources, and follow-up searches should take this into account.


\begin{acknowledgments}
This research has made use of data or software obtained from the
Gravitational Wave Open Science Center
(gwosc.org)~\cite{LIGOScientific:2019lzm,LIGOScientific:2025snk}, a
service of the LIGO Scientific Collaboration, the Virgo Collaboration,
and KAGRA. This material is based upon work supported by NSF's LIGO
Laboratory which is a major facility fully funded by the National
Science Foundation. The LIGO document number of this article is
LIGO-P2600518. The authors are grateful to Deepali Agarwal for
valuable discussions on this topic and for sharing her expertise on
the ASAF search, and to Federico De Lillo for his careful reading of
the manuscript and constructive feedback. Stavros Venikoudis is
supported by a grant from the Fonds pour la formation \`a la
Recherche dans l'Industrie et dans l'Agriculture (FRIA), awarded by
the Fonds de la Recherche Scientifique (F.R.S.-FNRS), Belgium. The
data analysis in this work was performed using
\texttt{NumPy}~\cite{vanderWalt}, \texttt{SciPy}~\cite{Virtanen}, and
\texttt{HEALPy}~\cite{Zonca2019}. All figures were produced with
\texttt{Matplotlib}~\cite{Hunter}. The sky pixelization uses the
HEALPix framework~\cite{Gorski:2004by}.
\end{acknowledgments}

\appendix

\section{Derivation of the CW cross-correlated power}
\label{app:cw_response}
The phase of a signal with frequency drift $\dot{f}$ is
$\Phi(t) = 2\pi(f_0 t + \dot{f}t^2/2)$. Over an SFT of duration $T$,
the drift adds a phase $\pi\dot{f}T^2$ to that of a monochromatic
signal. The largest drift comes from the rotation of the Earth. Its
centripetal acceleration, $a \simeq 0.03$~m\,s$^{-2}$, gives
$\dot{f}/f_0 \leq a/c \simeq 10^{-10}$~s$^{-1}$. Keeping the added
phase below $\pi/2$ requires
$T \lesssim 3\times10^{3}\,\mathrm{s}\,\sqrt{500\,\mathrm{Hz}/f_0}$.
Spin-down gives a weaker bound. The SFTs of $T = 192$~s used here
(Sec.~\ref{sec:csd}) satisfy both.

Within an SFT, the signal is therefore monochromatic at the
instantaneous frequency $f_0(t)$. The SFT of
Eq.~\eqref{eq:h_detector} then consists of two lines at $\pm f_0(t)$,
\begin{equation}
\begin{split}
  \tilde{h}_{I_i}(t;f)
  &= \frac{A_+F^+_{I_i} - iA_\times F^\times_{I_i}}{2}
     e^{i[\Phi_{I_i}(t){-}2\pi f t]}
     \delta_T(f{-}f_0(t))\\
  &\;+ \frac{A_+F^+_{I_i} + iA_\times F^\times_{I_i}}{2}
     e^{-i[\Phi_{I_i}(t){+}2\pi f t]}
     \delta_T(f{+}f_0(t))\,,
\end{split}
  \label{eq:sft_cw}
\end{equation}
where
\begin{equation}
  \delta_T(f)
  = \int_{-T/2}^{T/2} e^{-2\pi i f t'}\,\mathrm{d}t'
  = \frac{\sin(\pi f T)}{\pi f}
  \label{eq:deltaT}
\end{equation}
is the finite-time Dirac delta function~\cite{Allen:1997ad}, with
width $\sim 1/T$ and height $\delta_T(0) = T$. Multiplying the positive-frequency lines of the two detectors gives
Eq.~\eqref{eq:signal_crosspower}. The negative-frequency lines give
its complex conjugate at $-f_0(t)$. The initial phase, the intrinsic
phase evolution, and the reference phase $e^{-2\pi i f t}$ are common
to both detectors and cancel. Only the differential phase remains,
\begin{equation}
  \Delta\Phi_I(t)
  = \Phi_{I_2}(t) - \Phi_{I_1}(t)
  = 2\pi f\,
    \frac{\hat{n}\cdot\Delta\vec{x}^I(t)}{c}\,.
  \label{eq:dphi}
\end{equation}
Spin-down enters Eq.~\eqref{eq:dphi} only through the drift of
$f_0(t)$ times the light travel time across the baseline. Over an
O2-length run, this is below $2\times10^{-3}$~rad even at the
spin-down limit of Sec.~\ref{sec:heff}, and we neglect it.

\section{Closed form and modulation depth}
\label{app:closedform}
Equations~\eqref{eq:Fplus} and~\eqref{eq:Fcross} combine into
$F_i^+ + iF_i^\times = z_i\,e^{2i\psi}$, with $z_i = a_i + ib_i$, so
the polarization angle enters only as a phase. The two polarization
products are then
\begin{align}
  F_1^+F_2^+ + F_1^\times F_2^\times
    &= \mathrm{Re}\bigl(z_1\bar{z}_2\bigr) = A_t\,,\\
  F_1^+F_2^+ - F_1^\times F_2^\times
    &= \mathrm{Re}\bigl(z_1 z_2\,e^{4i\psi}\bigr)\,,
\end{align}
which are Eqs.~\eqref{eq:idsum} and~\eqref{eq:iddiff}, with
$C^{(1)}_t + iC^{(2)}_t = z_1z_2$. We write
$\mathrm{Re}\,G_t = \mathcal{A}_+^2F_1^+F_2^+
+ \mathcal{A}_\times^2F_1^\times F_2^\times$ as
$(\mathcal{A}_+^2+\mathcal{A}_\times^2)/2$ times the sum of the two
products plus $(\mathcal{A}_+^2-\mathcal{A}_\times^2)/2$ times their
difference. Inserting this in Eq.~\eqref{eq:mu2} with
$\gamma_t = A_t$ gives
\begin{equation}
  \mu^2
  = \frac{\mathcal{A}_+^2+\mathcal{A}_\times^2}{2}
  + \frac{\mathcal{A}_+^2-\mathcal{A}_\times^2}{2}\,
    \mathrm{Re}\!\left(e^{4i\psi}\,
    \frac{C^{(1)} + iC^{(2)}}{D}\right),
\end{equation}
which is Eq.~\eqref{eq:theorem}, since
$C^{(1)} + iC^{(2)} = \varepsilon D\,e^{i\delta}$ by
Eq.~\eqref{eq:epsdelta}. With uniform weights the time sums involve
only $a_i$ and $b_i$, so $\varepsilon$ and $\delta$ do not depend on
frequency. The bound $\varepsilon \leq 1$ ensures
$\mu^2 \geq \mathcal{A}_\times^2 \geq 0$, which keeps $\sqrt{\mu^2}$ in
Eq.~\eqref{eq:h_chain} real. 

With uniform weights, $\varepsilon$ depends on declination alone.
Each term of the time sums depends on sidereal time and right
ascension only through their difference. A change of right ascension
therefore shifts the sum over a full day and leaves it unchanged. At the
celestial poles the rotation of the Earth turns both detectors about
the line of sight. The phase of $z_1z_2$ then advances uniformly over
the day, and $\varepsilon = 0$. The modulation is largest on the two
rings at $\mathrm{Dec} \simeq \pm19.5^{\circ}$.

\bibliography{citations}
\end{document}